\documentclass[10pt]{article}

\usepackage{amsmath,amssymb,amsthm}
\usepackage{amsxtra}
\usepackage{amsfonts}
\usepackage{amssymb}
\usepackage{url}
\usepackage{hyperref}
\usepackage{esint}
\usepackage{eucal}
\usepackage{mathrsfs}
\usepackage{cite}
\usepackage{graphicx,graphics}

\newcommand\sect[1] {\ref{sect:#1}}
\newcommand\labsect[1] {\label{sect:#1}}

\newcommand\eq[1] {(\ref{#1})}
\newcommand{\bfm}[1]{\mbox{\boldmath ${#1}$}}

\newcommand{\nonum}{\nonumber \\}
\newcommand{\beqa}{\begin{eqnarray}}
\newcommand{\eeqa}[1]{\label{#1}\end{eqnarray}}
\newcommand{\beq}{\begin{equation}}
\newcommand{\eeq}[1]{\label{#1}\end{equation}}
\newcommand{\Grad}{\nabla}
\newcommand{\Div}{\nabla \cdot}
\newcommand{\Curl}{\nabla \times}

\newcommand{\Tr}{\mathop{\rm Tr}\nolimits}

\newcommand{\Md}{\partial}

\newcommand{\Ga}{\alpha}

\newcommand{\Gd}{\delta}
\newcommand{\Ge}{\epsilon}

\newcommand{\Gf}{\phi}

\newcommand{\Gk}{\kappa}

\newcommand{\Gn}{\eta}
\newcommand{\Gm}{\mu}

\newcommand{\Gt}{\theta}
\newcommand{\Gvt}{\vartheta}
\newcommand{\Gr}{\rho}

\newcommand{\Gs}{\sigma}

\newcommand{\Gj}{\tau}

\newcommand{\Go}{\omega}

\newcommand{\Gz}{\zeta}

\newcommand{\GO}{\Omega}

\newcommand{\BGa}{\bfm\alpha}
\newcommand{\BGb}{\bfm\beta}

\newcommand{\BGe}{\bfm\epsilon}
\newcommand{\BGve}{\bfm\varepsilon}

\newcommand{\BGn}{\bfm\eta}
\newcommand{\BGm}{\bfm\mu}

\newcommand{\BGs}{\bfm\sigma}

\newcommand{\BGG}{\bfm\Gamma}
\newcommand{\BGL}{\bfm\Lambda}

\newcommand{\BGY}{\bfm\Psi}

\newcommand{\CE}{{\cal E}}

\newcommand{\CI}{{\cal I}}
\newcommand{\CJ}{{\cal J}}
\newcommand{\CK}{{\cal K}}

\newcommand{\CM}{{\cal M}}

\newcommand{\CT}{{\cal T}}
\newcommand{\CU}{{\cal U}}

\newcommand{\BCC}{{\bfm{\cal C}}}
\newcommand{\BCD}{{\bfm{\cal D}}}

\newcommand{\BCQ}{{\bfm{\cal Q}}}
\newcommand{\BCR}{{\bfm{\cal R}}}
\newcommand{\BCS}{{\bfm{\cal S}}}

\newcommand{\bpm}{\begin{pmatrix}}
\newcommand{\epm}{\end{pmatrix}}

\def\Ba{{\bf a}}
\def\Bb{{\bf b}}

\def\Bd{{\bf d}}
\def\Be{{\bf e}}
\def\Bf{{\bf f}}
\def\Bg{{\bf g}}
\def\Bh{{\bf h}}

\def\Bj{{\bf j}}
\def\Bk{{\bf k}}

\def\Bm{{\bf m}}
\def\Bn{{\bf n}}

\def\Bp{{\bf p}}
\def\Bq{{\bf q}}
\def\Br{{\bf r}}
\def\Bs{{\bf s}}

\def\Bu{{\bf u}}
\def\Bv{{\bf v}}
\def\Bw{{\bf w}}
\def\Bx{{\bf x}}
\def\By{{\bf y}}

\def\BB{{\bf B}}
\def\BC{{\bf C}}
\def\BD{{\bf D}}
\def\BE{{\bf E}}
\def\BF{{\bf F}}
\def\BG{{\bf G}}

\def\BI{{\bf I}}
\def\BJ{{\bf J}}
\def\BK{{\bf K}}
\def\BL{{\bf L}}

\def\BP{{\bf P}}
\def\BQ{{\bf Q}}
\def\BR{{\bf R}}
\def\BS{{\bf S}}
\def\BT{{\bf T}}
\def\BU{{\bf U}}
\def\BV{{\bf V}}

\def\BY{{\bf Y}}
\def\BZ{{\bf Z}}

\usepackage{graphics}
\usepackage{amssymb}

\title{A unifying perspective on linear continuum equations prevalent in science. Part I: Canonical forms for static, steady, and quasistatic equations}
\author{}
\date{}
\begin{document}
\maketitle
\vskip -.5cm
\centerline{\large Graeme W. Milton}
\centerline{Department of Mathematics, University of Utah, USA -- milton@math.utah.edu.}
\vskip 1.cm
\begin{abstract}
  Following some past advances, 
  we reformulate a large class of linear continuum science
  equations in the format
  of the extended abstract theory of composites so that we can apply this theory to
  better understand and efficiently solve those physical equations.
  Here in part  I  we elucidate the
  form for many static, steady, and quasistatic equations. 
\end{abstract}
\section{Introduction}
\setcounter{equation}{0}
\labsect{1}
We are interested in linear science equations that can be manipulated into the form
\beq \BJ=\BL\BE-\Bs,\quad \BGG_1\BE=\BE,\quad\BGG_1\BJ=0,
\eeq{ad1}
as encountered in the extended abstract theory of composites.
Until Part  V \cite{Milton:2020:UPLV}  we will consider these equations in a medium of infinite extent,
possibly, though not necessarily, periodic.
The first equation is called the constitutive law and it is typically taken to be local in space with points represented by $\Bx=(x_1,x_2,x_3)$, i.e., $\BJ(\Bx)=\BL(\Bx)\BE(\Bx)-\Bs(\Bx)$. The field $\Bs(\Bx)$ is the source term, while
$\BGG_1$ is a projection operator in Fourier space. Thus if $\BL$ or $\BGG_1$ act on a field $\BF$ to produce a field $\BG$ then we have, respectively,
that $\widehat{\BG}(\Bk)=\BGG_1(\Bk)\widehat{\BF}(\Bk)$ or that  $\BG(\Bx)=\BL(\Bx)\BF(\Bx)$ , in
which $\widehat{\BG}(\Bk)$ and $\widehat{\BF}(\Bk)$ are the Fourier components of $\BG$ and $\BF$, and $\Bk=(k_1,k_2,k_3)$ represents a point
in Fourier space. By multiplying the constitutive law by $\BGG_1$ the
solution to \eq{ad1} is easily found to be
\beq \BE=(\BGG_1\BL\BGG_1)^{-1}\BGG_1\Bs \eeq{ad1sol}
where the inverse is to be taken on the space $\CE$ onto which $\BGG_1$
projects. In Part VI \cite{Milton:2020:UPLVI} we will see that the solution can be written in various
alternative forms that are useful for generating rapidly converging series
expansions for the solution.

The prototypical example is the conductivity equation (that we will revisit later):
\beq \Bj'(\Bx)=\BGs(\Bx)\Be(\Bx)-\Bs(\Bx),\quad \BGG_1\Be=\Be,\quad\BGG_1\Bj'=0, \quad\text{with}\quad \BGG_1=\Grad(\Grad^2)^{-1}\Grad\cdot,
\eeq{pt1}
where $\BGs(\Bx)$ is the conductivity tensor, while $\Div\Bs$, $\Bj=\Bj'+\Bs$, and $\Be$ are the current source, current,
and electric field, and $(\Grad^2)^{-1}$ is the inverse Laplacian
(there is obviously considerable flexibility in the choice of $\Bs(\Bx)$, the
only constraints being square integrability and that
$\Div\Bs$ equals the current source).
An interesting twist is that we write the equations in this form,
rather than in the more conventional form involving $\Bj$ directly, but this is exactly
what we need to keep the left hand side of the constitutive law divergence free.
As current is conserved, $\Div\Bj=\Div\Bs$, implying $\Div\Bj'=0$, which is clearly equivalent to
$\BGG_1\Bj'=0$. To show that $\Be=\Grad(\Grad^2)^{-1}\Grad\cdot\Be$, we let $V$ be the solution of Poisson's equation
$\Grad^2V=-\Div\Be$ (with $V(\Bx)\to 0$ as $\Bx\to\infty$), i.e. $V=-(\Grad^2)^{-1}\Div\Be$.
Then integrating this gives $\Be=-\Grad V=\Grad(\Grad^2)^{-1}\Div\Be=\BGG_1\Be$.
These steps are much easier done in Fourier space, where $\BGG_1(\Bk)=\Bk(\Bk\cdot\Bk)^{-1}\Bk^T$.

In this Part  I  we cast  a multitude of static, steady, and quasistatic
equations in this form. In Part II \cite{Milton:2020:UPLII} we continue with
time harmonic science equations.
In Part  III  \cite{Milton:2020:UPLIII} we express in the desired form a host of dynamic equations in stationary media; dynamic equations where the material is moving and where
the moduli vary with time. Part IV \cite{Milton:2020:UPLIV} includes
equations involving higher order gradients of the fields 
In Part  V \cite{Milton:2020:UPLV} we review how one can get
bounds on the spectrum of the relevant operator; and
reformulate the Gibiansky-Cherkaev transformation that we will also
discuss in this part, and obtain Stieltjes type integral representions
for the resolvent of non-selfadjoint operators. 
In Part  VI \cite{Milton:2020:UPLV} we review iterative methods for
solving for the fields $\BJ$ and $\BE$
given $\BL$, $\BGG_1$, and $\Bs$. These  iterative methods are based on
series expansions. 
In Part VII we address how boundary value problems and scattering problems can also be cast in this framework.
All parts are largely based on Chapter 2 of \cite{Milton:2002:TOC}  (reviewed in \cite{Allaire:2003:BRT, Sawicki:2003:RTC}),
and the book \cite{Milton:2016:ETC} (reviewed in \cite{Sharma:2017:BRE, Grabovsky:2018:BRE}), but extend the theory further.

The fields in \eq{ad1} are square integrable over all space, or if periodic, integrable over the unit cell of periodicity. We allow for nonperiodic fields
in periodic media provided they are square integrable over all space. At each point $\Bx=(x_1,x_2,x_3)$ the fields take values in a space $\CT$ of supertensors,
by which we mean a finite collection of scalars, vectors, and tensors. For most purposes
the tensorial nature of the fields is not important and, using an appropriate basis to provide a representation, we can just consider
$\BJ(\Bx)$, $\BE(\Bx)$ and $\Bs(\Bx)$ to be vector fields and $\BL(\Bx)$ and $\BGG_1(\Bk)$ to be matrix valued fields
in real and Fourier space respectively. Given two fields $\BP_1(\Bx)$ and
$\BP_2(\Bx)$ in this space of fields, we define the inner product of them
to be
\beq (\BP_1,\BP_2)=\int_{\mathbb{R}^3}(\BP_1(\Bx),\BP_2(\Bx))_{\CT}\,d\Bx,
\eeq{innp}
where $(\cdot,\cdot)_{\CT}$ is a suitable inner product on the space $\CT$
such that the projection $\BGG_1$ is selfadjoint with
respect to this inner product, and thus the space $\CE$ onto which
$\BGG_1$ projects is orthogonal to the space $\CJ$ onto which
$\BGG_2=\BI-\BGG_1$ projects. When we have periodic fields in periodic media the integral in \eq{innp} should be taken over the unit cell of periodicity.

The reason for keeping the subscript 1 on $\BGG_1$ is that if $\BL(\Bx)$ is nonsingular, \eq{ad1} can be rewritten in the
dual form
\beq  \BE=\BL^{-1}\BJ+\BL^{-1}\Bs,\quad \BGG_2\BJ=\BJ,\quad\BGG_2\BE=0,
\eeq{aee}
where now $\BGG_2=\BI-\BGG_1$ plays the role that $\BGG_1$ played in \eq{ad1},
and $-\BL^{-1}\Bs$ plays the role that $\Bs$ played. 

We are only selectively reviewing the literature in this series of articles,
mainly because of its broad scope. The reader is encouraged to look at the papers and books referenced and when a subject appeals,
to delve further by following the articles and books referenced in those papers, or alternatively to seek papers or books that reference them. The exposition here is slanted by the author's
perspective, thereby accounting for the many references to his and his collaborators  work.

To make contact with conventional representations of physics equations, we follow Section 12.2 of \cite{Milton:2002:TOC}
and note that typically $\BGG_1(\Bk)$ (or its blocks, or the corresponding blocks in $\BGG_2(\Bk)$) has the factorization
\beq \BGG_1(\Bk)={\BD}(i\Bk)[\BF(\Bk)]^{-1}{\BD}(i\Bk)^\dagger, \quad \text{where}\quad \BF(\Bk)={\BD}(i\Bk)^\dagger{\BD}(i\Bk),
\eeq{addd}
in which ${\BD}(i\Bk)$ is a (scalar, vector, tensor, or supertensor) polynomial function of $i\Bk$ and the inverse in \eq{addd} is to be taken on the range of
${\BD}(i\Bk)^\dagger$, defined as the adjoint of ${\BD}(i\Bk)$.
The action of ${\BD}(i\Bk)^\dagger$ in Fourier space is equivalent in real space to the action of the differential operator
$\BD(\Grad)$ and we can rewrite \eq{ad1} as
\beq \BD(\Grad)^\dagger\BL\BD(\Grad)\BGY=\Bf, \eeq{conv}
that may be solved for the (possibly multicomponent) potential $\BGY$, given a source term $\Bf$.
We can identify $\Bf$ with $\BD(\Grad)^\dagger\Bs$, and the potential field $\BGY$ may be obtained from $\BE$
through their Fourier transforms:
\beq \widehat{\BGY}(\Bk)=[\BF(\Bk)]^{-1}{\BD}(i\Bk)^\dagger\widehat{\BE}(\Bk).
\eeq{conv1}
Conversely, given $\Bf$ and a solution $\BGY$ we have that
\beq \BE=\BD(\Grad)\BGY,\quad \BJ=\BL\BE-\Bs,
\eeq{conv2}
provided $\Bs$ is chosen so that $\Bf=\BD(\Grad)^\dagger\Bs$. Note that the source $\Bs$ is not uniquely determined, as reflected in the
equations \eq{ad1} -- given a field $\BJ_0$ such that $\BGG_1\BJ_0=0$ we can add it to $\Bs$ and subtract it from $\BJ$,
without disturbing $\BE$ or $\BGY$. We can move back and forth between the two different formulations as we please, provided
$\BGG_1(\Bk)$ has the factorization \eq{addd}. It is not always simpler to introduce potentials. For instance, the
symmetric second order stress field $\BGs$, satisfying $\Div\BGs=0$ is the double
curl of the Beltrami stress tensor \cite{Beltrami:1892:OSN}, still a symmetric second order field.

The representation \eq{ad1} has the advantages:
\begin{itemize}
\item it unifies a large class of linear science equations, allowing
  general tools
  for solving them to be applied to, or developed for,
  any equation in the class. Conversely scientists developing theory for
  one type of equation should see if their theory applies to the whole class,
  or has already been derived in that context. The situation is quite
  reminiscent of scientists developing results for the conductivity equation,
  in the context of different physics problems, without realizing that
  essentially they were all solving the same problem, not being aware of 
  the commonality. For this reason different names have
  been attached to the same basic equations, such as the different names
  for the Clausius-Mossotti approximation, and for Bruggeman's effective
  medium theory \cite{Landauer:1978:ECI};
\item that problems sharing the same $\BGG_1(\Bk)$, or some of the
  blocks of $\BGG_1(\Bk)$, may have unexpected deeper connections. If $\BL(\Bx)$
  shares the same block diagonal structure as $\BGG_1(\Bk)$
  (perhaps after making some manipulations, such as those of
  Milgrom and Shtrikman \cite{Milgrom:1989:LRT}: see also Chapter 6
  of \cite{Milton:2002:TOC} and references therein), then the
  equations decouple into a subset of equations. Conversely, the off
  diagonal blocks of $\BL(\Bx)$ can be viewed as coupling this
  subset of equations together;
\item that the transition to its dual
  form \eq{aee} is automatic;
\item that   the form of $\BL$ often suggests additional
  ``bianisotropic type \cite{Serdyukov:2001:EBAM}'' couplings in the time
  harmonic wave equations at constant frequency, or for thermal conductivity equations in the Laplace domain;
\item that for these wave equations in lossy media, and for thermal or particle diffusion with complex frequencies, or more generally for problems
  with a non Hermitian tensor $\BL$ a host of variational minimization 
  principles naturally follow from it
  \cite{Cherkaev:1994:VPC, Milton:1990:CSP, Fannjiang:1994:CED,
    Norris:1997:LTB, Milton:2009:MVP, Milton:2010:MVP, Carini:2015:VFL,
    Milton:2017:BCP}
  that are not easy to extend to equations written in the form \eq{conv};
  
\item that the fields all lie in the same space, so that $\BL$ and its adjoint $\BL^\dagger$ act on the same space, making it sensible to add or subtract them,
  as required for the just mentioned manipulations;
\item that, as will be seen in Part  V, one can solve the equations using rapidly convergent series expansions;
  \item that, importantly, the space in which we solve \eq{ad1} is just the space of square integrable fields. Thus one can take the viewpoint that
Sobolev spaces can be dispensed with in the context of solving these linear science equations.
Gone are the derivatives of the fields and potentials. Instead one is left with the projection $\BGG_1(\Bk)$. One may assert that sophisticated mathematical
analyses are needed to establish existence and uniqueness of solutions. However, I believe that they are best addressed within the framework of \eq{ad1}:
see also Part  V;
\item that the formulation allows one to implement ideas developed in the theory of exact relations for 
composites \cite{Grabovsky:1998:EREa,Grabovsky:2000:ERE} (see also Chapter 17 of
\cite{Milton:2002:TOC} and the book \cite{Grabovsky:2016:CMM}) to obtain universal (geometry independent) exact identities satisfied by the infinite body
Green's function for inhomogeneous media when the material tensor $\BL(\Bx)$ takes values on certain manifolds $\CM$ in tensor space, i.e., $\BL(\Bx)\in\CM$
for all $\Bx$. These also lead to a flood
of new conservation laws called boundary field equalities \cite{Milton:2019:ERG}.
\end{itemize}

It also has disadvantages:
\begin{itemize}

\item  in real space the equations are nonlocal, not just involving differential operators;
\item the source term $\Bs$ may be less localized, particularly when the physical sources are $\Div\Bs$. For instance, in dielectrics, $\Div\Bs$ represents the
  charge density, so for just two well separated charges there must be a
  flux of $\Bs$ between them. It may be necessary to require that the net chargebe zero to ensure square integrability of the fields.
  This is the case in two dimensions (corresponding to three dimensional line charges);
\item  the original equations typically involve fewer variables (the potentials) rather than the fields directly;
\item  in some cases the representation is a lot more cumbersome, and less physically transparent.

\end{itemize}

It is not always the case that $\BF(\Bk)$ in \eq{addd} is a scalar or proportional to the identity matrix (examples being
the time harmonic electromagnetic equations and Midlin plate equations).
Nevertheless, in all cases I have encountered
$\BGG_1$ is block diagonal with block diagonal elements of the form ${\BG}(i\Bk)/g(k^2)$ (or ${\BG}(i\Bk,\Go)/g(k^2,\Go^2)$ for time
dependent problems) where
${\BG}(i\Bk)$ and $g(k^2)$ are polynomials in $i\Bk$ and $k^2$ respectively. Thus, the action of ${\BG}(i\Bk)$ is
equivalent to the action of $\BG(\Grad)$ in real space, and in Fourier space one only needs to compute the action of $1/f(k^2)$.
That many equations of science can be represented in the form \eq{conv} has been noted, for example, by Strang \cite{Strang:1986:IAMB, Strang:1988:FEE} [see also chapter 2
of \cite{Strang:2007:CSE}]. Here and in Parts II, III, and IV  we show that equations which can be represented in the form \eq{ad1} are ubiquitous in science.

Clearly \eq{conv} has the solution
\beq \BGY=\underline{\BR}\Bf,\quad\text{where}\quad \underline{\BR}=[\BD(\Grad)^\dagger\BL\BD(\Grad)]^{-1}. 
\eeq{conv3}
In the same way that the fundamental solution (Green's function) for $\BGY$
is obtained by taking a delta function source $\Bf(\Bx)=\Bf_0\Gd(\Bx-\Bx_0)$,
so too is the fundamental solution for $\BE$ in \eq{ad1} obtained
by taking a delta function source $\Bs(\Bx)=\Bs_0\Gd(\Bx-\Bx_0)$
in \eq{ad1sol}.

In many problems of interest, including many time harmonic wave equations including the time harmonic Schr{\"o}dinger equation, $\BD(\Grad)^\dagger\BL\BD(\Grad)$
can be written in the form
\beq \BD(\Grad)^\dagger\BL\BD(\Grad)=\underline{z}\BI-\BD(\Grad)^\dagger\underline{\BB}\BD(\Grad).
\eeq{conv4}
Thus, for example, for typical wave equations $\underline{z}=\Go^2$, where $\Go$ is the frequency, and for the Schr{\"o}dinger equation $\underline{z}=E$,
where $E$ is the energy. However, we emphasize that for
time harmonic wave equations in inhomogeneous media, the moduli and hence
$\underline{\BB}$ also typically depend on
$\Go$ due to the dispersive nature of the materials.

Substituting \eq{conv4} back in \eq{conv3} gives
\beq \underline{\BR}=[\underline{z}\BI-\BD(\Grad)^\dagger\underline{\BB}\BD(\Grad)]^{-1}, \eeq{conv5}
and we have the resolvent problem that many scientists focus on solving, rather than \eq{ad1}. Here $\underline{\BB}$ acts in a different space
than where $\Bf$ and $\BGY$ live. We will come back to studying resolvents in Parts  V and VI.

One can treat the case where $\BGG_1(\Bk)$ has blocks that factor in a similar way. Thus, for example, suppose that $\BGG_1(\Bk)$ has the factorization
\beq \BGG_1(\Bk)=\bpm {\BD}_1(i\Bk)[\BF_1(\Bk)]^{-1}{\BD_1}(i\Bk)^\dagger
& 0 \\ 0 & {\BD}_2(i\Bk)[\BF_2(\Bk)]^{-1}{\BD_2}(i\Bk)^\dagger \epm,
\eeq{blf1}
where $\BF_1(\Bk)={\BD_1}(i\Bk)^\dagger{\BD_1}(i\Bk)$ and 
$\BF_2(\Bk)={\BD_2}(i\Bk)^\dagger{\BD_2}(i\Bk)$.
Then an equivalent problem is to solve
\beq \bpm {\BD_1}(\Grad)^\dagger
& 0 \\ 0 & {\BD_2}(\Grad)^\dagger \epm
\BL\bpm {\BD_1}(\Grad)
& 0 \\ 0 & {\BD_2}(\Grad)\epm\bpm \BGY_1 \\ \BGY_2 \epm
= \bpm \Bf_1 \\ \Bf_2 \epm.
\eeq{blf2}

While the examples given here and in Parts II, III, and IV, which are
mostly drawn from Chapter 2 of \cite{Milton:2002:TOC} and Chapter 1 of \cite{Milton:2016:ETC},
are extensive, they are by no means exhaustive. Huge families of equations that can be expressed in this form are given
in Section 12.2 of \cite{Milton:2002:TOC} and in \cite{Milton:2013:SIG}
(with an addendum in \cite{Milton:2015:ATS}).

We emphasize that the form \eq{ad1} only requires one to specify $\BL(\Bx)$ and $\BGG_1(\Bk)$ or $\BGG_2(\Bk)=\BI-\BGG_1(\Bk)$.
From a mathematical viewpoint is not even necessary to identify what $\BJ$, $\BE$ and $\Bs$ represent physically,
although this can be helpful to give an intuitive understanding of the solutions and their physical significance.

The way of writing the equations in the desired form is not unique. Even so,
a good way of doing this is sometimes by no means obvious.
One learns as one manipulates more and more equations into the appropriate form.
A general procedure that always works is hard to identify. It helps to
find $\BGG_1$ from the derivatives of the potentials, although to
complicate matters potentials can also occur on the left hand side
of the constitutive relation. With enough
struggling one typically meets with success, if necessary by introducing
appropriate  auxiliary fields. 
The underlying reason for this success is not so clear except in the cases
where the equations result from some
energy minimization associated with thermodynamic considerations. A deeper
understanding is clearly needed. It is also puzzling that a few equations
do not seem to be amenable to be put in this form. Examples include
perturbations to the Navier Stokes equations with viscous terms,
and, even neglecting shear viscosity,
perturbations to the magnetohydrodynamics of the ionosphere
\cite{Kelley:2009:EIP} and perturbations to the equations governing atmospheric
behavior \cite{Panofsky:1970:AAB}. 

In some cases the operator $\BL$ has a nontrivial null space, and this
brings into question the solution \eq{ad1sol}. To avoid this
one can often shift $\BL(\Bx)$ by a
multiple $c$ of a ``null-$\BT$ operator''
$\BT_{nl}(\Bx)$ (acting locally in real space), defined to have the
property that
\beq \BGG_1\BT_{nl}\BGG_1=0, \eeq{nl-1}
that then has an associated quadratic form (possibly zero) that is
a ``null-Lagrangian''.
Clearly the equations \eq{ad1} still hold, with $\BE(\Bx)$ unchanged
and $\BJ(\Bx)$ replaced by $\BJ(\Bx)+c\BT_{nl}\BE(\Bx)$
if we replace $\BL(\Bx)$ with  $\BL(\Bx)+c\BT_{nl}(\Bx)$.
We will come back to discussing
``null-$\BT$ operators'' in Section 3  of Part V. In a few other
cases $\BL$ may contain $\infty$ (or $\infty$'s) on its diagonal.
If one can remove any degeneracy of $\BL(\Bx)$, we can consider the
dual problem \eq{aee} and then, if desired, try to shift  $\BL^{-1}(\Bx)$
by a multiple
of a ``null-$\BT$ operator'' $\widetilde{\BT}_{nl}(\Bx)$ satisfying
$\BGG_2\widetilde{\BT}_{nl}\BGG_2=0$ to remove its degeneracy. 

The relevant physical fields are progressively defined, and we do not typically remind the reader of their
definitions in subsequent equations. We emphasize that when space derivatives occur in the fields
entering the constitutive relations, one can easily rewrite them
in a form where no space  derivatives occur in the fields entering the
constitutive relations. Rather such differential constraints
on the fields are embodied in $\BGG_1(\Bk)$ and the formula for it can be deduced from the differential constraints on the fields that are explicit in
Parts I, II, III, and IV.

To avoid taking unnecessary transposes, we let $\Div$ act on the first index of a field, and the action of $\Grad$ produces a field, the first index of
which is associated with $\Grad$. References are provided to those equations that might not be familiar to the general reader.

\section{Electrical conductivity and similar equations, both statics and quasistatics}
\setcounter{equation}{0}
\labsect{3}

These are the simplest of all the equations and take the form
\beq \Bj'=\BL\Be-\Bs,\quad \Div\Bj'=0,\quad \Be=-\Grad V, \eeq{15.1}
where, for electrical conductivity, $\Div\Bs$, $\Bj=\Bj' + \Bs$, $\Be$, and $V$ are the source of electrical current,
electric current, electric field, and potential, that are complex in quasistatics.
For fluid flow in porous rocks (Darcy's law),
thermally conducting materials (Fick's law), and in materials where
particle diffusion occurs, the same equations apply with $\Div\Bs$, $\Bj=\Bj' +\Bs$, $\Be$, being replaced by the appropriate fields. Respectively,
$\Div\Bs$ is a source of fluid in
a porous medium, a source of heat flux, or a source of particle flux;
$\Bj=\Bj' + \Bs$ gets replaced by the macroscopic fluid velocity field $\Bv$, heat flux $\Bq$, and particle current; $\Be$
is replaced by the pressure gradient, temperature gradient, particle
concentration gradient; $V$ by the pressure, temperature, or
particle concentration.

Bubbly flow in a nonviscous fluid, with the bubbles
rising or moving along a pipe all with the same velocity is described by
the conductivity equations when one is in a frame of reference moving with
the bubbles \cite{Smereka:1991:TDP}. The bubbles carry along with them some fluid when they rise and hence have a virtual mass. For an ensemble of moving bubbles the
effective virtual mass density corresponds to the effective conductivity.
Sedimenting particles that all sediment at the same rate can be treated
similarly.

For dielectrics, after expressing the free charge density $\Gr_f$ as $\Gr_f=\Div\Bs_\Gr$, the electrostatic equations take the form \eq{15.1}
with $\Bj'$ replaced by $\Bd'=\Bd-\Bs_\Gr$, where $\Bd$ is the electric displacement field and $\Bs$ is replaced by $\Bs_d=\Bs_\Gr-\Bp$
where  $\Bp$ is the permanent electric polarization plus the polarization due to the pyroelectric effect when the temperature is changed.

Now for all these equations we have:
\beq \BL(\Bx)=\BGs(\Bx), \quad \BGG_1(\Bk)=\frac{\Bk\otimes\Bk}{k^2}=\frac{{\BD(i\Bk)}{\BD(i\Bk)}^\dagger}{k^2},\quad\text{with  }\BD(\Grad)=\Grad, \eeq{x1}
where for conductivity materials, dielectrics, porous rocks, thermally conducting materials, and in materials where
particle diffusion occurs $\BGs(\Bx)$ is the conductivity tensor, electric permittivity $\BGve(\Bx)$,
rock permeability $\Bk(\Bx)$ divided by the dynamic viscosity of the
fluid $\Gm$, thermal conductivity $\BK(\Bx)$, and diffusivity $\BD(\Bx)$ respectively.

For magnetism, after expressing the free current $\Bj_f$ as $\Bj_f=\Curl\Bs_j$, the equations become
\beq \Bh'=\BL\Bb-\Bs, \quad \Curl\Bh'=0,\quad \Div\Bb=0, \eeq{15.1+}
in which $\Bs=\Bm+\Bs_j$, where $\Bm$ is the permanent magnetization, $\Bh=\Bh'+\Bs_j$ is the magnetic field, and $\Bb$ the magnetic induction.
We then have
\beq \BL=[\BGm(\Bx)]^{-1},\quad \BGG_1(\Bk)=\BI-\frac{\Bk\otimes\Bk}{k^2}, \eeq{15.2+}
where $\BGm(\Bx)$ is the magnetic permeability tensor. These equations are dual to those in \eq{15.1} and \eq{x1} and we may reexpress them as
\beq \Bb=\BGm\Bh'+\BGm\Bs, \quad  \BGG_2(\Bk)\Bh'=\Bh',\quad \BGG_2(\Bk)\Bb=0, \quad \BGG_2(\Bk)=\BI- \BGG_1(\Bk)=\frac{\Bk\otimes\Bk}{k^2},
\eeq{15.2++}
which, after swapping the indices in $\BGG_1$ and $\BGG_2$, is then of the same
form as \eq{15.1} and \eq{x1}.

The equations for stellar radiative transfer of
energy also take this form,
where over a small frequency interval
from $\Go$ to $\Go+\Ge$, $V$ would be the energy density $U_\Go$ in that
frequency interval, and $\Bj$ is then the radiative flux in this frequency
interval. Replacing $\BGs$ is $D_\Go\BI$ where the diffusion
constant $D_\Go$ is given by
$D_\Go=\tfrac{1}{3}c\ell_\Go$ in which $c$ is the speed of light and
$\ell_\Go$ is the mean free path of photons having frequency $\Go$.
One can also write $\ell_\Go=1/(\Gr\Gk_\Go)$ where $\Gr$ is the
density and $\Gk_\Go$ is the opacity coefficient at frequency $\Go$
\cite{Pols:2009:SSE}.
The energy density can be connected to the temperature $T$
through the Planck distribution:
\beq U_\Go=\frac{8\pi h \Go^3}{c^3(e^{h\Go/k_BT}-1)}, \eeq{srt1}
where $h$ is Planck's constant and $k_B$ is Boltzman's constant.
Using $\Grad U_\Go=[\Md U_\Go/\Md T]\Grad T$ and integrating over
frequencies, gives again a conductivity (Fick's law) equation
with $\Grad V$ being the temperature gradient
and $\Bj$ now being the total radiative flux, with $\BGs$ being
related to $T^3$ times the reciprocal of the
Rosseland mean opacity $\Gk$ \cite{Pols:2009:SSE}:
\beq \BGs=\frac{4acT^3}{\Gk\Gr},\quad
\frac{1}{\Gk}=
\frac{\int_0^\infty[\Md U_\Go/\Md T]/\Gk_v\,d\Go}
{\int_0^\infty [\Md U_\Go/\Md T]\,d\Go}
=\frac{1}{4aT^3}\infty[\Md U_\Go/\Md T]/\Gk_v\,d\Go,
\eeq{srt1aa}
where $a$ is the radiation constant, $a=8\pi^5k_B^4/(15h^3c^3)$
As an approximation one may take the opacity coefficient $\Gk_\Go$
to be independent of $\Go$, giving $\Gk_\Go=\Gk$. Then $\BGs$
scales as $T^3$.
If one has radial symmetry so that $\Grad=\widehat\Br\Md/\Md r$, where
$\widehat\Br$ is the radial unit vector, then by integrating over $r$
we see that the total energy flux scales as $T^4$ which is the
Stefan-Boltzmann law. 

Additionally, the two dimensional conductivity equations hold for torsion
for small twists around the $x_3$-axis provided one assumes the elasticity
tensor $\BCC(\Bx)$ only depends on $x_1$ and $x_2$ and is invariant with
respect to reflection about the $x_3=0$ plane. Thus all elements
$C_{ijk\ell}$ of the elasticity tensor are zero if the set of indices
$\{i,j,k,\ell\}$ contains an odd number of 3's. Then, in
Saint Venant's theory of torsion \cite{Atkin:1980:ITE, Fung:1965:FSM},
the components of the displacement field $\Bu(\Bx)$ take the form
\beq u_1=-\Gj x_3 x_2, \quad u_2=+\Gj x_3 x_1,\quad u_3=u(x_1,x_2),
\eeq{tor1}
 where $\Gj$ is the amplitude of the twist, and $u(x_1,x_2)$ is the warping
function. The linear elasticity equations imply
\beq \BGs_3'\equiv\underbrace{\bpm \Gs_{13}' \\ \Gs_{23}' \epm}_{\BJ}
=\BL \underbrace{\bpm \Md u/\Md x_1 \\ \Md  u/\Md x_2 \epm}_{\BE}
- \underbrace{\left[\BL \bpm \Gj x_2 \\ -\Gj x_1 \epm+\Bs'\right]}_{\Bs},
\quad \frac{\Md \Gs_{13}'}{\Md x_1}+\frac{\Md \Gs_{23}'}{\Md x_2}=0,
\eeq{tor2}
where $\Div\Bs'(x_1, x_2)$ is a shear forcing in the vertical direction
(not the torque force which is applied to surfaces of constant $x_3$
at the ends of the cylinder, that may have arbitrary cross section),
and the shear vector field $\BGs_3=\BGs_3'+\Bs$ has
components $\Gs_{13}(\Bx)=\Gs_{31}(\Bx)$ and $\Gs_{23}(\Bx)=\Gs_{32}(\Bx)$
that are the only nonzero components of the stress field $\BGs$. We have
\beq \BL=\bpm C_{1313} & C_{1323} \\ C_{1323} & C_{2323} \epm,\quad
\BGG_1=\Bk\otimes\Bk/k^2.
\eeq{tor3}
The square integrability of $\Bs$, assuming $\Bs'$ is appropriately chosen
is ensured if $\BL(\Bx)$ is zero beyond some distance from the $x_3$-axis.
While the linear elasticity
assumption that $\Bu(\Bx)$ is small is not satisfied at sufficiently
large $|x_3|$, this does not matter as what is important is that
the stresses and strains be small enough: the equation \eq{tor3} is
independent of $x_3$ which reflects the fact that the physics is the same,
modulo a rotation, when we change $x_3$. 

Antiplane elasticity is a special case of torsion where $\Gj=0$, but
the shear forcing $\Bs'$ can be nonzero. Then the displacement field $\Bu(\Bx)$
is aligned with the $x_3$-direction, i.e. $u_1(\Bx)=u_2(\Bx)=0$.

\section{Thermoelectricity and magnetoelectricity}
\setcounter{equation}{0}
\labsect{4}
For thermoelectricity and magnetoelectricity
the  equations take the form \cite{Callen:1960:TIPa}:
\beqa \bpm -\Bj_N' \\ \Bj_U' \epm & = & \BL\bpm \Grad(\Gm/T) \\ \Grad(1/T) \epm -\bpm -\Bs_N \\ \Bs_U \epm,\quad \Div\Bj'_N=\Div\Bj_U'=0, \nonum
  \bpm \Bd' \\ \Bb \epm & = & \BL\bpm \Be \\ \Bh' \epm -\bpm \Bs_d \\ \Bs_b \epm ,\quad \Div\Bd'=\Div\Bb=0,\quad \Curl\Be=\Curl\Bh'=0.
    \eeqa{15.2}
For thermoelectricity, $\Div\Bs_N$ and  $-\Div\Bs_U$ are the sources of electrons and energy,
$\Bj_N=\Bj_N'+\Bs_N$ and $\Bj_U=\Bj_U'+\Bs_U$ are the current densities of the number of electrons and energy; $\Gm$ and $T$ are the electrochemical potential per particle and the temperature. For magnetoelectricity,
the fields are those in the previous section, including for $\Bs_d$
    contributions coming from the free charge density
    and permanent polarization; and for $\Bs_b$  contributions coming from the free current density and permanent magnetization - also note that
     $\Bh=\Bh'+\Bs_j$ is the magnetic field, where $\Curl\Bs_j=\Bj_f$ is the free current.

The tensor $\BL$ in these, and other coupled equations of the conductivity type, takes the form
\beq \BL(\Bx)=\bpm \BL_{11} & \BL_{12} \\ \BL_{21} & \BL_{22} \epm,\quad \BGG_1(\Bk)= \frac{1}{k^2}\bpm \Bk\otimes\Bk & 0 \\ 0 & \Bk\otimes\Bk\epm
=\frac{1}{k^2}{\BD(i\Bk)}{\BD(i\Bk)}^\dagger,\quad\text{with  }  \BD(\Grad)=\bpm \Grad &  0 \\ 0 & \Grad \epm,
\eeq{x2}
where generally $\BL_{21}=(\BL_{12})^{T}$. Effects associated with the
thermoelectric coupling are the Seebeck effect, where a temperature gradient causes a current, and Peltier effect, where an electric current causes a temperature gradient,
\cite{Callen:1960:TIPa}. Since 1976
thermoelectrics have been used to convert heat into power in  spacecraft. 

Generally, ordinary materials do not have a significant magnetoelectric coupling.
An exception is $\textrm{Cr}_2\textrm{O}_3$. 
We will come across other sources of magnetoelectric couplings shortly.
Multiferroics have the interesting property that the magnetic source term
(magnetization) is linked with the electric source term (electric polarization).


\section{Quasistatic complex dielectric equations}
\setcounter{equation}{0}
\labsect{5}
Associated with \eq{15.1} are quasistatic equations where the fields and $\BL(\Bx)$ are complex.
In quasistatics the physical fields are the real parts of $e^{-i\Go  t}\Bj$, $e^{-i\Go  t}\Be$, $e^{-i\Go  t}V$, and  $e^{-i\Go  t}\Bs$
respectively. The quasistatic electric and dielectric equations get mixed, as time varying displacement fields and free charge densities
give current fields and free current fields:
\beq \Bj=-i\Go\Bd, \quad \Bs_j=-i\Go\Bs_\Gr,\quad \Bj'=-i\Go\Bd'. \eeq{jereln}
We assume that the permanent electric polarization fields and magnetic polarization fields are absent, though one could easily
include in the source field time varying electric polarization fields resulting from the pyroelectric effect with a
temperature oscillating at frequency $\Go$. Due to this mixing, the complex conductivity (admittivity) tensor $\BGs$ is related to the complex permittivity tensor
$\BGve$ via $\BGs=-i\Go\BGve$, where $\Go$ is the frequency.

To understand the origin of the quasistatic equations in electromagnetism note that the
time harmonic electromagnetic equations imply
\beq \Curl\Be=i\Go\Bb,\quad \Curl\Bh-\Bj_f=-i\Go\Bd, \eeq{15.1aa}
where $\Bb$ and $\Bh$ are the magnetic field and magnetizing field. 
If the gradients in $\Be$ and $\Bh$ are very large, or if $\Go$ is such that the structure or inhomogeneities in material properties
are much smaller than the wavelengths or attenuation lengths, then one can neglect the right hand sides of the above equations
that act like source terms: this is the quasistatic approximation. Thus in the quasistatic approximation there is a decoupling
into the quasistatics dielectric equations and the quasistatic magnetic equations. The quasistatics equations are valid
in the context of dielectrics, magnetic materials, flow in porous materials, and in materials where
particle diffusion occurs, but not for thermal conduction. For flow in porous media, the imaginary part of the permeability $\Bk$ is due to the dynamic viscosity arising from
the viscous interaction of the fluid with the pore surface, related to what is called the tortuosity reflecting its increase when the fluid passes through
a more tortuous pore geometry \cite{Johnson:1987:TDP}). The form of the equation for dynamic thermal conduction is presented in Section 2.1 of Part  III.

The equations \eq{15.1} can be manipulated into the extended Cherkaev-Gibiansky form
    \beq \bpm \Be\\ -i\Bd' \epm=\BL \bpm -i\Bd'\\ \Be \epm -\Bs_0,\quad \Div\Bd'=0,\quad \Be=-\Grad V,
    \eeq{15.3}
    where $\Be$ and $\Bd'$ are complex ($\Bd'=\Bd-\Bs_\Gr$, where $\Bd$ is the displacement field, should not be confused with the real part of $\Bd$),
    but $\BL$ is Hermitian. This is a modification of the form originally proposed by Cherkaev and Gibiansky \cite{Cherkaev:1994:VPC}
    with source terms added \cite{Milton:2010:MVP}, and with no splittings of $\Be$, $\Bd'$
    and $\Bs$ into their real and imaginary
    parts (which is not necessary, as we will see in Section 7 of part  V). They are now coupled equations with
    \beqa \BL& = & \bpm
[\BGve'']^{-1} & i[\BGve'']^{-1} \BGve' \\
-i\BGve'[\BGve'']^{-1} & \BGve''+ \BGve'[\BGve'']^{-1} \BGve' \epm,\quad \Bs_0=\bpm -[\BGve'']^{-1}\Bs \\ (\BI+i\BGve'[\BGve'']^{-1})\Bs \epm, \nonum
\BGG_1(\Bk)& = &\frac{1}{k^2}\bpm k^2\BI-\Bk\otimes\Bk & 0 \\ 0 & \Bk\otimes\Bk\epm
=\frac{1}{k^2}{\BD(i\Bk)}{\BD(i\Bk)}^\dagger,\quad\text{with  }  \BD(\Grad)=\bpm \Grad \times &  0 \\ 0 & \Grad \epm,
\eeqa{x3}
where the additional source term is that introduced in \cite{Milton:2010:MVP} and
where $\BGve'$ and $\BGve''$ are the real and imaginary parts of the complex electrical permittivity tensor $\BGve(\Bx)=\BGve'(\Bx)+i\BGve''(\Bx)$.
The advantage of this reformulation is
that $\BL$ is Hermitian and is positive definite if and only if $\BGve''(\Bx)$ is positive definite and this allows one to use
minimization variation principles to obtain bounds
on the complex  effective permittivity tensor of a periodic composite \cite{Milton:1990:CSP}. The imaginary parts of the permittivity tensor,
or equivalently the real part of the conductivity
tensor accounts for energy loss due to resistive heating and thus $\BGve''(\Bx)$ is positive semidefinite,
and typically positive definite if $\BGve'(\Bx)$ is not positive definite.
By multiplying \eq{15.1}, and hence $\BL$ by $e^{i\Gvt}$ where $\Gvt>0$ is small, we can convert to an equivalent problem where
 $\BGve''(\Bx)$ is positive definite. 

We remark that with respect to bounding the effective tensors of composites (or the associated problem of bounding the Dirichlet to Neumann map governing
the response of inhomogeneous bodies -- see chapters 3, 4, and 5 in \cite{Milton:2016:ETC}),
that beside variational principles, a parallel approach for multicomponent media has been to obtain bounds
using the analytic properties developed in \cite{Golden:1983:BEP, Bergman:1978:DCC, Golden:1985:BEP} of the effective moduli
as a function of the component moduli. The analytic properties when $\BL(\Bx)$ is piecewise constant taking $N$ values
$\BL_1$, $\BL_2$, ...$\BL_N$ 
corresponding to an $N$-phase medium extend to all the equations encountered in this paper: see Chapter 18 of \cite{Milton:2002:TOC},
and Chapters 3 and 4 in  \cite{Milton:2016:ETC}. There is no need to rederive them for individual equations.
These analytic properties are generally Herglotz type properties, with {\it(i)} the effective
tensor (or Dirichlet to Neumann map ) being a homogeneous analytic function of degree 1 of the component moduli, or component tensors,
{\it(ii)} being analytic when appropriate component moduli have positive imaginary parts, or component tensors have positive definite imaginary parts, 
and then {\it(iii)} with the imaginary part of the effective tensor (or the imaginary part of the Dirichlet to Neumann operator) being positive definite.
This reflects the causal nature of the response, and typically the fact that the
composite absorbs energy when the component moduli absorb energy. In fact, the original bounds on the complex effective electrical permittivity of a composite of two isotropic phases
(with sufficient symmetry that the effective conductivity tensor is isotropic) were first derived using the
analytic properties \cite{Bergman:1980:ESM, Milton:1980:BCD}.
More recently they have been improved
\cite{Kern:2020:RCE} using variational methods based on the formulation \eq{x3}. Generally, when there are more that two component moduli, the variational method
usually yields tighter bounds, applies to media with continuously varying moduli (not just two phase composites) and moreover is easier to implement.
Some exceptions are that it remains to be seen if variational methods can recover bounds correlating the values the effective electrical permittivity takes at
more than two frequencies \cite{Milton:1981:BTO, Clark:1995:OBC} or recover bounds on the
response of lossy media at a specific time \cite{Mattei:2016:BRV, Mattei:2016:BRL}.

As briefly mentioned in Section \sect{2}, the customarily accepted notion that conduction at low enough frequencies is governed by the equations
\beq \Bj(\Bx,t)=\BGs_0(\Bx)\Be(\Bx,t)+\BGve_0(\Bx)\frac{\Md}{d t}\Be(\Bx,t),
\eeq{exc1}
is not always correct. In the frequency domain this corresponds to a complex conductivity $\BGs(\Bx)=\BGs_0(\Bx)-i\Go\BGve_0(\Bx)$ or a complex
electrical permittivity $\BGve(\Bx)=\BGve_0(\Bx)+i\BGs_0(\Bx)/\Go$. Now consider, for example, a checkerboard array of conducting square cylinders with complex
electrical permittivity  $\BGve_1(\Go)=\BGve_1+i\BGs_1/\Go$, alternating with a second nonconducting material with a real
electrical permittivity $\BGve_2$. In the transverse plane the electrical permittivity is $\BGve^*(\Go)=\sqrt{\BGve_1(\Go)\BGve_2(\Go)}$ (see \cite{Dykhne:1970:CTD})
having a real part that diverges as the $\Go\to 0$ and correspondingly the current $\Bj(\Bx,t)$ is a nonlocal convolution in time of $\Be(\Bx,t)$. A similar divergence
of the real part of the complex permittivity constant as $\Go\to 0$ occurs in porous rocks containing conducting salt water: see \cite{Stroud:1986:AMD}
and references therein.
For most purposes, one can get rid of such nonlocalities by approximating the response in the frequency domain by a (perhaps matrix or operator valued)
rational function. Then the response can be modeled using a
multifield approach \cite{Mariano:2005:CAM}
by introducing hidden variables such as, for example, collections of
oscillators. This works 
even for the Maxwell equations \cite{Raman:2010:PBS}.

With $\Go=ip$, $p$ being real, $e^{-i\Go t}=e^{pt}$ is also real and hence so too are all the fields, implying that $\BGve(\Bx)$ (and $\BGs(\Bx)$) is real
in this case. It follows from this and from the analyticity, implied by causality, of $\BGve(\Bx,\Go)$ as a function of $\Go$ in the domain where $\Go$ has positive
imaginary part, that
\beq \BGve(\Bx,\Go)=\overline{\BGve(\Bx,-\overline{\Go})}, \eeq{recip}
where the overline denotes complex conjugation.

In plasmonic materials, such as metals such as silver or gold, at infrared wavelengths the real part of $\BGve(\Bx)$ can be negative with a small imaginary
part. The sea of electrons in these metals oscillate with respect to the applied electric field and when the oscillation is near $180^\circ$ out of phase
one gets an $\BGve(\Bx)$ with these properties. It can be the case, such as in laminated composites where these plasmonic materials are layered with normal
dielectric materials, that the real part of the effective tensor
$\BGe_*$, being the homogenized $\BGve(\Bx)$ can have both negative and positive eigenvalues with a small imaginary part so that in the effective medium
the conductivity equation \eq{15.1} becomes essentially a wave equation. Such media have the power of subwavelength resolution
\cite{Jacob:2006:OHF,Salandrino:2006:FFS}. 
Also, surprisingly, they can convert an incoming plane wave into beams (both going in the same direction as the incident wave or in an opposite direction)
originating from the boundary of a cylindrical hole in the material \cite{Milton:2013:SEH}.

In nonpassive materials the real part of $\BGs(\Bx,\Go)$, or equivalently the imaginary part of $\BGve(\Bx,\Go)$, can be negative.
An example is a tunnel diode where the current through it is not a monotonic function of the voltage across it (see, for example, Section 7.5
in \cite{Omar:1975:ESS}).
As a consequence the tangent moduli become negative and the diode becomes unstable, producing an oscillatory response.
It can become a ``battery'' producing energy at the oscillation frequency. At optical frequencies, organic dyes and other gain media used in lasers
can have a negative imaginary part of $\BGve(\Bx,\Go)$, again acting as an energy source now driven by the pumping of the material into
a higher energy state. 
\section{Magnetotransport, the Faraday effect, and convective diffusion in an incompressible stationary convective flow}
\setcounter{equation}{0}
\labsect{6}
    In the presence of a magnetic field the electrical conductivity equations \eq{15.1} still hold true but the conductivity tensor
    depends on the magnetic field and is not symmetric due to a breaking of time reversal symmetry. The nonsymmetric part is tied to
    the Hall effect where, for example, in an isotropic conductor
    a magnetic field induces an additional electric field perpendicular to
    both the magnetic field and the electrical current. Hall effect sensors
    provide a useful tool for measuring magnetic field (which is how the
    compass in your cell phone works). 

    Similarly, the effective permittivity tensor is not symmetric in the presence of
    a magnetic field, causing a phenomena known as the Faraday effect where there is a nonreciprocal rotation of the plane of polarization of electromagnetic waves (to be distinguished from the reciprocal rotation associated with
    optical activity).
    Additionally the magnetic permeability tensor is not symmetric at microwave frequencies
    in ferrites with an external biasing magnetic field, and also leads to a
    much stronger Faraday effect \cite{Hogan:1952:FFE},
    An important application is in the construction of gyrators, circuit elements that ensure one way propagation of signals: see, for example,
    \cite{Hogan:1952:FFE, Pozar:2012:ME}.

    Likewise, the convective diffusion equations governing temperature or particle
    concentration diffusion in an incompressible stationary convective flow can also be manipulated into the form \eq{15.1} with a
    ``conductivity tensor'' $\BGs$ having an antisymmetric part $\BGs_a$ determined by the fluid flow field velocity $\Bv$ \cite{Fannjiang:1994:CED}:
    one chooses a
    $\BGs_a(\Bx)$ such that $\Div\BGs_a=\Bv$, which is possible if the fluid is incompressible, i.e., $\Div\Bv=0$. 
    Then the conductivity equation implies the  stationary heat equation with heat source $\Bs(\Bx)$:
    \beq 0=\Div\Bj=\Div(\BGs_s+\BGs_a)-\Bs=\Div\BGs_s\Grad T-\Bs + \Bv\cdot\Grad T+\Tr(\BGs_a\Grad\Grad T), \eeq{he}
    where the last term vanishes because $\BGs_a$ is antisymmetric, while $\Grad\Grad T$ is symmetric.

    Manipulations based on the Cherkaev-Gibiansky transformation
    \cite{Cherkaev:1994:VPC} result in an equivalent formulation \cite{Milton:1990:CSP} (see also \cite{Fannjiang:1994:CED, Norris:1997:LTB}) 
    taking the form \eq{x3}. Specifically, $\Bd'$ is replaced by $\Bj'$, or the heat flux minus $\Bs$, or the particle current minus $\Bs$,
    and $\Be$ remains the same for
    magnetotransport but is replaced by the temperature or particle concentration gradients for convective diffusion, and
    $\BL$ given by
\beq \BL=\bpm \BGs_s^{-1} & -\BGs_s^{-1}\BGs_a \cr \BGs_a\BGs_s^{-1} & \BGs
_s-\BGs_a\BGs_s^{-1}\BGs_a \epm, \quad \BGG_1(\Bk)=\frac{1}{k^2}\bpm k^2\BI-\Bk\otimes\Bk & 0 \\ 0 & \Bk\otimes\Bk\epm,
\eeq{x3a}
where $\BGs_s$ and $\BGs_a$ are the symmetric and antisymmetric parts of $\BGs=\BGs_s+\BGs_a$. For convective diffusion $\BGs_s$ can be
identified with the thermal conductivity tensor $\BK(\Bx)$ or diffusivity tensor $\BD(\Bx)$.
The tensor $\BL$ is real and symmetric and is positive definite if and only if $\BGs_s(\Bx)$ is positive definite. This form of the equations leads to tight bounds on magnetotransport in
composites \cite{Briane:2007:BSF}, thus proving its worth. Similarly, the variational principle met with
success when applied to convective diffusion in periodic stationary incompressible flows \cite{Fannjiang:1994:CED}. Of course, more generally, one can do such manipulations when $\BL(\Bx)$ is not Hermitian, and then
the resulting tensor entering the constitutive law will be Hermitian, and additionally will be positive definite if and and only if the
Hermitian part of $\BL$ is positive definite. 

\section{Linear elasticity, with or without prestress, for both statics and quasistatics}
\setcounter{equation}{0}
\labsect{7}
We consider linear elasticity in the presence of body forces $\Bf$ that we write as $\Bf=-\Div\Bs_f$. These body forces may include
gravity or forces generated by electric charges (when there is an electric field) or electric dipoles (when there is a gradient in the electric field)
embedded in the material (such materials are called electrets \cite{Kressmann:1996:SCE}) or due to magnetic fields if the material is magnetic.
Then the elasticity equations take the form \cite{Atkin:1980:ITE, Fung:1965:FSM}:
\beq \BGs'=\BL(\Bx)\Grad\Bu -\Bs,\quad \Div\BGs=0,\quad \BGe=[\Grad\Bu+(\Grad\Bu)^T]/2,
\eeq{15.4}
where $\BGs=\BGs'+\Bs_f$, $\BGe$, $\Bu$, are the stress, strain, and displacement field, while $\Bs=\Bp_e+\Bs_f$ where $\Bp_e$ is the ``free elastic
polarization'' due to prestress (that may perhaps arise
during manufacturing), swelling due to humidity, or thermal expansion.
Dislocations can also be handled by adding appropriate source terms
\cite{Berbenni:2014:NSA, Bertin:2018:FBF}.

The bending of
thin plates is also described by \eq{15.4}, as will be discussed further in
Section 4.1 of Part  IV .
Prestresses are not time
harmonic, but $\Bs$ may be time harmonic, with the quasistatic equations applying, if there is a time harmonic humidity or
expansion due to temperature fluctuations, or time harmonic electric fields acting on an electret.

In all these circumstances we have
\beq \BL(\Bx)=\BCC(\Bx), \quad  \BGG_1(\Bk)=\underline{\BP}(\Bk),
\eeq{x4}
where $\BCC(\Bx)$ is the fourth order elasticity tensor (which for quasistatics has an imaginary part reflecting the viscoelastic contribution)
and
$\underline{\BP}(\Bk)$ is the projection operator that projects onto matrices of the form $\Ba\otimes\Bk+\Bk\otimes\Ba$. Its action on a second order tensor
$\BC$ is given by
\beq \underline{\BP}(\Bk)\BC=
\left[(\BC\Bk)\otimes\Bk+\Bk\otimes(\BC\Bk)\right]/k^2-\left[(\Bk\cdot\BC\Bk)\Bk\otimes\Bk\right]/k^4,
\eeq{x4a}
and it has elements
\beq \{\underline{\BP}(\Bk)\}_{ij\ell m}=\tfrac{1}{2}\left(k_i\Gd_{j\ell}k_m+k_i\Gd_{jm}k_\ell+k_j\Gd_{i\ell}k_m+k_j\Gd_{im}k_\ell\right)/k^2-k_ik_jk_\ell k_m/k^4.
\eeq{x5}

If one is interested in the response to
temperature fluctuations accounting for the entropy, one should consider the full equations of linear thermoelasticity, which we introduce now.
\section{Static linear thermoelasticity and static poroelasticity}
\setcounter{equation}{0}
\labsect{8}

We now have the equations \cite{Norris:1992:CBP}:
\beq \begin{pmatrix}\BGe \\ \Gz \end{pmatrix}=\BL\begin{pmatrix}\BGs'\\ \Gt\end{pmatrix}-\bpm \Bs_e \\ \Bs_f \epm,\quad  \BGe=[\Grad\Bu+(\Grad\Bu)^T]/2,
\quad \Div\Bs'=0,
\eeq{15.5}
where $\Gt$, the temperature increment, is independent of $\Bx$.
For thermoelasticity the strain source term $\Bs_e$ can include all the source terms mentioned in the previous section (that were associated with stress
but equivalently can be associated with strain), except the thermal expansion contributions that we are accounting for separately here,
and $\Bs_f$ can be a source of heat content,
while $\Gt$ and $\Gz$ are increase in temperature and entropy per unit volume over that when $\BGs=0$ and $\Gt=0$. The poroelastic
equations are the homogenized equations for a porous elastic material containing fluid, in particular porous rocks containing oil
or salt water. The stain source term $\Bs_e$ can include the same sources mentioned in the previous section now
contributing to the strain in  the solid phase, while $\Bs_f$ can be a source of fluid, $\BGe$ is the strain in
the solid phase, $\BGs$ is the confining stress,
and $\Gz$ and $\Gt$ are increment of the fluid content measuring the net flow of fluid in or out of a region and the negative of the pore fluid pressure.
Here $\Gt$ is constant while $\Gz$ is not subject to any differential constraints.

Associated with \eq{15.5} we have that
\beq \BL=\begin{pmatrix}\BCS(\Bx) & \BGa(\Bx) \\ \BGa(\Bx) & c(\Bx)/T_0 \end{pmatrix}
,\quad \BGG_1(\Bk)=\bpm \BI-\underline{\BP}(\Bk) & 0 \\ 0 & 0 \epm\quad
\text{for}\quad \Bk\ne 0,
\eeq{x6}
where for thermoelasticity $\BCS(\Bx)$ is the fourth order compliance tensor (the inverse of the elasticity tensor $\BCC(\Bx)$),
$\BGa(\Bx)$ is the symmetric second order tensor of thermal expansion,
$c(\Bx)$ is the specific heat per unit volume at constant stress (this specific heat measures
the amount of additional heat energy or, more precisely, entropy
that is stored in the material when the temperature is
increased at constant stress). 
If are we are not interested in keeping
account of the heat energy then \eq{x6} reduces to the elasticity equations \eq{x4} with an additional source term $\Bs=-\BGa(\Bx)\Gt$,
where $\Gt$ is the temperature increase. For poroelasticity $\BCS(\Bx)$ represents the compliance tensor of the drained frame,
$\BGa(\Bx)$ represents the tensor of fluid pressure induced expansion at constant confining stress, and the constant
$c(\Bx)$ relates the increment of fluid content to the fluid pressure, again at constant confining stress.

\section{Couplings between static elastic, electric, and magnetic fields}
\setcounter{equation}{0}
\labsect{9}

In many media, unless symmetry considerations prevent it, there are couplings between static elastic, electric, and
magnetic fields that are more pronounced in some materials than others. The equations take the form:
\beq \bpm \BGe \\ \Bd' \\ \Bb \epm  =  \BL\bpm \BGs' \\ -\Grad V \\ \Bh' \epm -\bpm \Bs_e \\ \Bs_d \\ \Bs_b \epm
,\quad \Div\Bd'=\Div\Bb=0, \quad\Div\BGs'=0,\quad \BGe=[\Bu+(\Bu)^T]/2,\quad
\Curl\Bh'=0,
\eeq{15.6}
in which $\Bs_e$, $\Bs_d$, and $\Bs_b$ represent the elastic, electric, and magnetic sources, as detailed previously, $\BGs'$ is related to the stress $\BGs$
as in Section \sect{7}, $\Bd=\Bd'+\Bs_\Gr$ is the electric displacement field, while $\Bh=\Bh'-\Bs_j$ is the magnetic field, where $\Bs_j$ is such that
$\Curl\Bs_j=\Bj_f$, in which $\Bj_f$ is the free current. We have
\beq \BL(\Bx)=\begin{pmatrix}\BCS(\Bx) & \BCD(\Bx) & \BCQ(\Bx) \\
 \BCD^T(\Bx) & \BGve(\Bx) & \BGb(\Bx) \\
 \BCQ^T(\Bx) & \BGb^T(\Bx) & \BGm(\Bx)\end{pmatrix}, \quad \BGG_1(\Bk)=\bpm \BI-\underline{\BP}(\Bk) & 0 & 0\\ 0 & \Bk\otimes\Bk & 0 \\
0 & 0 & \Bk\otimes\Bk\epm,
\eeq{x6a}
where $\BCS(\Bx)$ is the compliance tensor when electric and magnetic fields are zero, $\BGve(\Bx)$ is the permittivity tensor when there is no strain
and no magnetic field, $\BGm(\Bx)$ is the magnetic permeability when there is no strain
and no electric field, $\BCD(\Bx)$ is a piezoelectric coupling term, $\BCQ(\Bx)$ is a piezomagnetic coupling term, and  $\BGb(\Bx)$ is a
magnetoelectric coupling term. Canonical examples of materials with  piezoelectric and  piezomagnetic couplings are $\textrm{BaTiO}_3$ and 
$\textrm{CoFe}_2\textrm{O}_4$ respectively. These can be used as actuators,
for ultrasonic sound generation, and for hydrophone applications
\cite{Avellaneda:1998:CPP}.  
Some of these coupling terms could be zero. Interestingly, one can have a composite where $\BGb(\Bx)=0$ everywhere,
yet the effective tensor has a nonzero magnetoelectric coupling term
(see, for example, \cite{Avellaneda:1994:MEP})--- this being an example of product properties \cite{Albers:1973:PPC}.
Physically an electric field in the piezoelectric phase creates a stress that stresses the piezomagnetic phase, producing a magnetic field.
Magnetoelectric couplings are also achieved in
multiferroics \cite{Spaldin:2019:AMM}, through couplings of the
electric Maxwell stress and magnetic Maxwell stress \cite{Liu:2013:GUM}
(discussed later in Section \sect{10.b})
and by depositing charges on the interface between two layers
of materials that differ in their magnetic properties \cite{Tan:2020:SRM}.

\section{Quasistatic viscoelasticity equations}
\setcounter{equation}{0}
\labsect{10}
These are given by \eq{15.4} with the moduli, fields and source term complex.
The actual fields and source terms are then the real parts of $e^{-i\Go t}$
times the complex fields and source term. The imaginary parts of the
moduli account for viscous losses \cite{Fung:1965:FSM}.

Besides the
elasticity setting the quasistatic viscoelasticity equations are valid
for small oscillations in a possibly compressible fluid containing
possibly compressible particles, assuming the  small oscillations
are superimposed on a base state where the fluid is at rest. The shear
modulus is then $\Gm=-i\Go\Gn$ where $\Gn$ is the shear viscosity of the fluid
(the real part of  $\Gm$ is zero because the fluid, assuming it is Newtonian,
cannot support static shearing). The complex bulk modulus $\Gk$ can have both a
real and imaginary part, with the imaginary part corresponding to a bulk
viscosity. The homogenized equations take the same form as for the
homogenized elasticity with an  effective  shear viscosity $\Gn_*$ and an
effective complex bulk modulus $\Gk_*$. Back in 1905 Einstein had calculated the effective shear viscosity of a suspension of rigid spheres in a fluid
\cite{Einstein:1905:BM}.

The complex equations
may be expressed in the modified Cherkaev-Gibiansky form \cite{Cherkaev:1994:VPC},
as gave \eq{x3}, to obtain from \eq{15.4} the equivalent formulation:
\beq \bpm \BGe\\ -i\BGs' \epm=\BL \bpm -i\BGs'\\ \BGe \epm -\Bs_0,\quad \Div\BGs'=0,\quad  \BGe=[\Grad\Bu+(\Grad\Bu)^T]/2,
\eeq{15.7}
as first presented in \cite{Cherkaev:1994:VPC}, then extended to include source
terms \cite{Milton:2010:MVP}, and finally with splittings of the fields into their real and imaginary parts avoided
as outlined in Section 7 of part  V . These developments entirely parallel those for the complex conductivity equation, and similar to
\eq{x3}, \eq{15.7} holds with $\Bu$ and $\BGs'$ being complex and
\beq \BL=\bpm
[\BCC'']^{-1} & i[\BCC'']^{-1} \BCC' \cr
-i\BCC'[\BCC'']^{-1} & \BCC''
+ \BCC'[\BCC'']^{-1} \BCC' \epm, \quad \BGG_1(\Bk)=\bpm \BI-\underline{\BP}(\Bk) & 0 \\ 0 & \underline{\BP}(\Bk) \epm,\quad
\Bs_0=\bpm -[\BCC'']^{-1}\Bs \\ (\BI+i\BCC'[\BCC'']^{-1})\Bs \epm.
  \eeq{x7}
This reformulation allows one to use minimization variational principles  \cite{Cherkaev:1994:VPC} to obtain bounds on the complex bulk and shear
moduli of viscoelastic composites \cite{Gibiansky:1993:EVM, Milton:1997:EVM}. These variational principles are appropriate when one is interested in bounds
on the response to oscillations at a constant frequency $\Go$, when the microstructure is much smaller than the wavelength.
Allowing for more general time dependencies, Carini and Mattei \cite{Carini:2015:VFL, Mattei:2017:BOP} split the time interval that the source has been
switched on into half and then apply Gibiansky-Cherkaev type transformations, like those leading to \eq{x3a}. Thus they obtain variational principles
and bounds for the viscoelastic response in the time domain that interlink  the responses in the first and second time periods.
They also mention a connection with the work of Tonti \cite{Tonti:1984:VFE},
but this misses the main point of transformations similar to \eq{x3} or \eq{x3a}: that one transforms to another problem of the type \eq{ad1},
and that, for composites, the effective tensor (or operator) of this problem is the transform of the effective tensor (or operator) of the
original problem. Otherwise, one may just as well minimize the integral of
the square of the equation, with possible weightings, and this is essentially what Tonti does.

\section{Steady viscous flow of an incompressible
liquid around an object}
\setcounter{equation}{0}
\labsect{10a}
The Oseen equations describe the steady flow of fluid around an object
moving with steady velocity $\BV$. The fluid is assumed to be viscous,
incompressible with $\BV$ small enough that the Reynolds number is
negligible. Far enough from the object the fluid can be considered to be at
rest. In the moving
frame of reference where the object is stationary the equations
\cite{Batchelor:2000:IFD} are
\beq -\Gr(\BV\cdot\Grad)\Bw=-\Grad P+\Gn\Grad^2\Bw +\Bf,\quad \Div\Bw=0,
\eeq{Ose1}
in which the total flow velocity in this moving frame of reference is $\Bw-\BV$,
$\Bf(\Bx)$ is a body force moving at the same speed as the object, and $\Gr$
and $\Gn$ are the fluid density and shear viscosity.
One has the boundary conditions that
$\Bw(\Bx)=\BU$ on the surface of the object and $\Bw(\Bx)\to 0$
as $|\Bx|\to\infty$. The equations become
\beq \bpm \BGs_s-P\BI \\ \\ \Div(\BGs_s-P\BI)\epm
=\BL\bpm \Grad\Bw \\ \Bw \epm,
\eeq{Ose2}
with
\beq \BL=\bpm \Gn\BGL_s +\infty\BGL_h & 0 \\
\Gr\BV\cdot & 0\epm,
\quad 
\BGG_1(\Bk)=\BZ(\Bk), \quad\text{where}\quad
\BZ(\Bk)\equiv\bpm i\Bk \\ 1 \epm\bpm -i\Bk & 1 \epm
=\frac{1}{1+k^2}\bpm \Bk\otimes\Bk & i\Bk \\
\\ -i\Bk & \BI \epm,
\eeq{Ose3}
in which $\BZ(\Bk)$ only acts on the first index of the matrix in
$(matrix,vector)$ fields,
$\BGL_s$ and  $\BGL_h$ are projections onto trace free symmetric
matrices, including $\BGs_s$, and matrices proportional to $\BI$, respectively, 
$\infty$ should be considered to be a large parameter that we 
let approach infinity. In this limit $\Div\Bw$ is forced to zero, corresponding
to the incompressibility of the flow,
while $P(\Bx)$ is unconstrained, except through the $\Grad P$ term
in \eq{Ose2}. Heat is
generated from the viscous flow and the corresponding temperature increase
will modify $\Gn$ as it is temperature dependent. Thus one
can expect that  $\Gn$ should depend on $\Bx$. One could use
an iterative procedure to solve the equations, where one
begins with a constant $\Gn$ and then at each iteration updates $\Gn(\Bx)$
according to the temperature distribution derived from
estimates of the heat generated obtained from the solution
for $\Bw(\Bx)$ at each iteration. Of course to obtain this
temperature distribution one will also need to take into
account the flow of heat and the heat capacity of the liquid.  

There are no analogous linear equations for compressible flows around
an object since the conservation of mass requirement that $\Div(\Gr\Bv)=0$
is a nonlinear constraint as $\Gr$ is related to the pressure through
the equation of state.  


\section{Steady viscous electron flow in graphene}
\setcounter{equation}{0}
The linearized
equations for  two-dimensional viscous electron flow, such as in a sheet of
graphene \cite{Bandurin:2016:NLR, Polini:2020:VEF}, are
\beq \Gs_0\Grad\Gf+D_\ell\Grad^2\Bj=\Bj, \quad \Div\Bj=0,
\eeq{gr1}
where $\Bj(\Bx)$ is the electrical current, $\Gs_0$ the diffusive
conductivity, and $D_\ell$ a diffusion
type coefficient representing the length over which the flow's momentum
diffuses, and $\Gf$ the electric potential. 
When $D_\ell=0$ this reduces to the standard conductivity
equation (Ohm's law).
These can be rewritten as
\beq \bpm \BQ \\ \Div\BQ \epm=\BL \bpm \Grad\Bj \\ \Bj \epm,
\quad \BL=\bpm (D_\ell/\Gs_0)\BGL_r+\infty\BGL_f & 0 \\
0 & 1/\Gs_0 \epm,
\eeq{gr2}
where $\BQ(\Bx)$ is a matrix valued flux, that takes the
form $\BQ=\Gf\BI+\BQ_f$ where $\BQ_f(\Bx)$ is trace free, though not
necessarily symmetric and $\BGL_h$ and $\BGL_f$ are the
projections onto matrices proportional to $\BI$ and trace free matrices,
respectively. Thus we have that $\BQ_f=\BGL_f\BQ$ and  $\BGL_h+\BGL_f=\CI$,
where $\CI$ is the fourth order identity tensor. Also $\infty$
represents a parameter that we should let approach infinity: this forces
$\Div\Bj=0$ and $\BQ_h=\BGL_h\BQ=\Gf$ represents a sort of pressure.
Accordingly, $\BGG_1=\BZ(\Bk)$ where $\BZ(\Bk)$ is given by \eq{Ose3}.

\section{Perturbations of magnetohydrostatic equations}
\setcounter{equation}{0}
\labsect{10.aa}
The magnetohydrostatic equations describe stationary states of a current
carrying, nonmagnetic conducting fluid such as a plasma. As such, they have
been important in describing fluorescent lights, astrophysical jets,
the solar corona,
the solar wind, and to the development of nuclear fusion in devices like
tokamaks. Assuming the fluid is perfectly conducting, highly compressible,
and light enough that one can neglect gravity, the equations take the simple form \cite{Bobbio:2000:EMF},
\beq \Div\Bb=0,\quad \Curl\Bb=\Gm_0\Bj, \quad
\Grad P=\Bj\times\Bb. \eeq{mhs0}
These are nonlinear due to the term $\Bj\times\Bb$ so we
look for equations satisfied by the first order perturbations.
We replace $\Bb$, $\Bj$, and $P$ with $\Bb+\Ge\Bb'$, $\Bj+\Ge\Bj'$, and
$P+\Ge P'$. Substituting these back in \eq{mhs0} and collecting
all terms of order $\Ge$ gives the equations
\beq \Div\Bb=0,\quad \Grad P=\Bj'\times\Bb+\Bj\times\Bb'
=-\Bb\times\Curl\Bb'/\Gm_0+\Bj\times\Bb',
\eeq{mhs0.a}
where $\Gm_0$ is the magnetic permeability of the vacuum. These imply
\beq \bpm P' \\ \Grad P' \\ \Bg' \epm
=\BL\bpm \Div\Bb' \\ \Bb'\\ \Curl\Bb' \epm,\quad \Curl\Bg'=0,
\eeq{mhs}
where $\Bg'$ is an auxillary field that the constitutive law with
\beq \BL=\bpm \infty & 0 & 0 \\ 0 & \BGn(\Bj) &  -\BGn(\Bb)/\Gm_0 \\ 0 & 0 & 0 \epm, \eeq{mhs0.b}
forces $\Bg'=0$: it is introduced to make the form of $\BGG_1(\Bk)$ clear.
Accordingly, we have
\beqa
\BGG_1(\Bk) & = & \bpm i\Bk^T \\ \BI \\ i\BGn(\Bk) \epm
[\Bk\times\Bk+\BI-\BGn(\Bk)\BGn(\Bk)]^{-1}\bpm -i\Bk & \BI & i\BGn(\Bk) \epm
\nonum
& = &\frac{1}{k^2+1}\bpm i\Bk^T \\ \BI \\ i\BGn(\Bk) \epm
\bpm -i\Bk & \BI & i\BGn(\Bk) \epm,
\eeqa{mhs1}
where the action of the second order tensor $\BGn(\Bk)$ on vector $\Ba$
gives $\BGn(\Bk)\Ba=\Bk\times\Ba$, and we have used the identity
$\BGn(\Bk)\BGn(\Bk)=\Bk\otimes\Bk-k^2\BI$.  As in the
previous section, the $\infty$ in $\BL(\Bx)$
should be considered to be a large parameter that we 
let approach infinity.  In this limit $\Div\Bb'$ is forced to zero,
while $P'(\Bx)$ is unconstrained, except through the $\Grad P$ term
in \eq{mhs}.


\section{Perturbations with Maxwell Stesses Present}
\setcounter{equation}{0}
\labsect{10.b}
The contribution to the stress from static electric fields, in the absence
of magnetic fields, is the Maxwell stress
\beq \BGs_e(\Bx)=\BGve(\Bx)[\Be(\Bx)\otimes\Be(\Bx)]-\tfrac{1}{2}\BGve(\Bx)[\Be(\Bx)\cdot\Be(\Bx)].
\eeq{max1}
Let us replace $\Be$ by $\Be+\Ge\Be'$ and correspondingly replace, to
first order in $\Ge$, the electric displacement field $\Bd$
with $\Bd+\Ge\Bd'$, the permanent polarization $\Bp$
by $\Bp+\Ge\Bp'$, the elastic displacement field $\Bu$ with $\Bu+\Ge\Bu'$,
the total stress $\BGs$ with $\BGs+\Ge\BGs'$, and the body force density
$\Bf$ with $\Bf+\Ge\Bf'$. Then, in the
absence of any magnetic fields, these perturbed fields are
related by
\beq \bpm \BGs' \\ \Div\BGs' \\ \Bd' \\ \Div\Bd' \epm=\BL
\bpm \Grad\Bu' \\ \Bu' \\ \Div\Be' \\ \Be'
\epm-
\bpm 0 \\ \Bf' \\ -\Bp' \\ 0 \epm.
\eeq{max2}
Note that $\Gr_c$, the density of charges embedded in the material,
does not appear in the source term for $\Bd'$
because we are only considering the
perturbed fields. We have 
\beqa \BL& = & \bpm \BCC & 0 & \BY_e(\Be) & 0 \\
 0 & \Be\otimes[\Grad \Gr_c -\Grad(\Div\Bp)] & 0 & (\Gr_c- \Div\Bp)\BI \\
               0 & (\Grad\Bp)^T+[(\Grad\BGve)\Be]^T & 0 & \BGve \\
               0 & (\Grad \Gr_c)^T & 0 & 0 \epm, \nonum
               \BGG_1 & = & \bpm \BZ(\Bk) & 0 \\ 0 & \BZ(\Bk) \epm,
               \eeqa{max3}
in which the action of the second order tensor $\BY_e(\Be)$ on $\Be'$
is given by
\beq \BY_e(\Be)\Be'=\BGve[\Be\otimes\Be'+\Be'\otimes\Be
-(\Be\cdot\Be')\BI],
\eeq{max4}
in which $\BZ(\Bk)$ is given by \eq{Ose3} and 
the first $\BZ(\Bk)$ in the $\BGG_1(\Bk)$ here
only acts on the first index of the matrix in $(matrix,vector)$ fields while
the second $\BZ(\Bk)$ in $\BGG_1(\Bk)$ acts on $(vector, scalar)$ fields. 
The terms $\Grad \Gr_c$, $\Grad(\Div\Bp)$ in \eq{max3} are due to the
movement of the embedded electrical charges and embedded dipoles associated
with the permanent polarization field, while the term $[(\Grad\BGve)\Be]^T$
is due to movement of the medium. 

Similar results hold in when one has static magnetic fields in the absence
of electric and current fields. The contribution to the stress from the magnetic
induction $\Bb$ is the Maxwell stress:
\beq \BGs_m(\Bx)
=[\BGm(\Bx)]^{-1}\Bb(\Bx)\otimes\Bb(\Bx)-\tfrac{1}{2}[\BGm(\Bx)]^{-1}[\Bb(\Bx)\cdot\Bb(\Bx)].
\eeq{max5}
We replace $\Bb$ by $\Bb+\Ge\Bb'$ and correspondingly replace, to
first order in $\Ge$, the magnetic field $\Bh$ with $\Bh+\Ge\Bh'$,
the permanent magnetization $\Bm$ with $\Bm+\Ge\Bm'$,
elastic displacement field $\Bu$ with $\Bu+\Ge\Bu'$,
the total stress $\BGs$ with $\BGs+\Ge\BGs'$, and the body force density
$\Bf$ with $\Bf+\Ge\Bf'$. Then, in the
absence of any electric fields and currents, the perturbed fields are
related by
\beq \bpm \BGs' \\ \Div\BGs' \\ \Bh' \epm=\BL
\bpm \Grad\Bu' \\ \Bu' \\ \Bb' \epm-
\bpm 0 \\ \Bf' \\ \Bm' \epm.
\eeq{max6}
We have 
\beq \BL=\bpm \BCC & 0 & \BY_m(\Bb)\\
               0 & \Bb\otimes[\Grad(\Div\Bm) & (\Div\Bm)\BI \\
               0 & (\Grad\Bm)^T+[(\Grad\BGm^{-1})\Bb]^T & \BGm^{-1} \epm,
               \quad
               \BGG_1 = \bpm \BZ(\Bk) & 0 \\ 0 & \BI-\Bk\otimes\Bk/k^2 \epm,
               \eeq{max7}
               in which the action of the second order
               tensor $\BY_m(\Bb)$ on $\Bb'$ is given by
\beq \BY_m(\Bb)\Bb'=\BGm^{-1}[\Bb\otimes\Bb'+\Bb'\otimes\Bb
-(\Bb'\cdot\Bb)\BI].
\eeq{max8}
Of course in the case where both electric fields and magnetic field
but no current fields are present, the perturbed equations take the form 
\beq
\bpm \BGs' \\ \Div\BGs' \\ \Bd' \\ \Div\Bd' \\ \Bh' \epm=\BL
\bpm \Grad\Bu' \\ \Bu' \\ \Div\Be' \\ \Be'\\ \Bb'
\epm-\bpm 0 \\ \Bf' \\ -\Bp' \\ 0 \\ \Bm' \epm,\quad
\BGG_1(\Bk)= \bpm \BZ(\Bk) & 0 & 0 \\ 0 & \BZ(\Bk) &0
\\ 0 & 0 & \BI-\Bk\otimes\Bk/k^2
\epm.
\eeq{max9}
In a composite a magnetic field can induce Maxwell stress that then
induces an electric field. This is again a product property,
that may account for the magnetic compass in birds \cite{Liu:2013:GUM}.


\section{Generating equations of the desired form from
  other linear and nonlinear equations}
\setcounter{equation}{0}
\labsect{2}
Equations of the form \eq{ad1} give rise to similar equations if $\BL(\Bx)$ and $\Bs$ are slightly perturbed. Replacing $\BJ$, $\BE$, $\BL$, and $\Bs$ with
$\BJ+\Ge\BJ'$, $\BE+\Ge\BE'$, $\BL+\Ge\BL'$, and $\Bs+\Ge\Bs'$ gives, to first order in $\Ge$,
\beq \BJ'=\BL\BE'+\BL'\BE-\Bs', \eeq{perturb}
which is essentially the same equation but with a new source term $\Bs'-\BL'\BE$. Thus the new source term is generally nonzero even if $\Bs=0$.

Small changes in the permittivity constant of materials can be due to
photoelasticity. Maxwell \cite{Maxwell:1851:EES}
found that the photoelastic effect, discovered by Brewster
\cite{Brewster:1833:EDL}, was linear in
the strain. The photoelastic effect can easily be seen if one looks at
layers of stressed transparent sticky tape on a glass surface or
stressed transparent plastic utensils between two crossed polarizing
sunglasses. It is used to detect cracks through the associated stress
causing a photoelastic effect.
The change $(\BGve^{-1})'$ to the inverse permittivity tensor $\BGve^{-1}$
takes the form
\beq (\BGve^{-1})'=\widetilde{\BCS}\BGe+\BCR[\Grad\Bu-(\Grad\Bu)^T]/2,
\eeq{pef}
where $\BGe=\tfrac{1}{2}[\Grad\Bu+(\Grad\Bu)^T]$
is the strain, the fourth order tensor $\widetilde{\BCS}$ is the photoelastic
tensor, while the fourth order tensor $\BCR$ is the rotoelastic
tensor \cite{Nelson:1972:BSA}. Besides the photoelastic effect there
is the acousto-optic effect, which also has many
applications: see the references in \cite{Smith:2017:EAO} where 
they show that the acousto-optic properties of layered structures
can be enhanced beyond those of the constituent materials.

Small nonsymmetric contributions to the conductivity $\BGs(\Bx)$,
permittivity $\BGve(\Bx)$ and permeability $\BGm(\Bx)$ can come from weak
magnetic fields due to the Hall effect, electric Faraday effect, and magnetic
Faraday effect, as discussed in Section \sect{6}.  Other changes to $\BL$ may come from variations in the
temperature, humidity, pressure, frequency (in quasistatics), or other
fields. An interesting example of where $\BL(\Bx)$ is perturbed by the
temperature is for thermoelectricity and this leads to the Thomson effect
where there is heating or cooling of a conductor carrying an electrical
current and having a temperature gradient along it \cite{Callen:1960:TIPa}.

Furthermore, many physical equations become an equation of a different type or separate into subsets of uncoupled equations
when the fields $\BJ(\Bx)$, $\BE(\Bx)$, $\BL(\Bx)$ and $\Bs(\Bx)$ are all independent of $\Bn\cdot\Bx$ for some vector $\Bn$, meaning that
they are constant in one direction.
Thus, for example, the conductivity or dielectric equation becomes a two dimensional equation of the pyroelectric type. To see this, we may take
$\Bn$ aligned with the $x_3$ axis. Then for $\Grad V$ to be independent of $x_3$ the potential $-V(x_1,x_2,x_3)$ must be of the form $e_3x_3-V'(x_1,x_2)$
where $e_3$ is constant and can be identified with the third component of the electric field $\Be$. The dielectric equations become
\beq \bpm d_1 \\ d_2 \epm=\bpm \Ge_{11} &  \Ge_{12} \\  \Ge_{21} &  \Ge_{22}\epm  \bpm e_1 \\ e_2 \epm+ \bpm \Ge_{13} \\  \Ge_{23} \epm e_3,
\quad \frac{\Md d_1}{\Md x_1}+\frac{\Md d_2}{\Md x_2}=0,\quad \frac{\Md e_1}{\Md x_2}-\frac{\Md e_2}{\Md x_1}=0,
\eeq{pyro}
which then have the form of two dimensional pyroelectric equations with $e_3$ playing the role of temperature (and $d_3$ not being subject to any differential
constraint). Similarly, the elasticity equations become a sort of  two dimensional piezoelectric equation
coupled with a scalar field, and for certain choices of moduli this can decouple into an  equation of the thermoelastic type and an equation
of the dielectric (conductivity) type, or into an equation of two dimensional elasticity and a two dimensional pyroelectric type:
see Sections 2.6 and 2.7 of \cite{Milton:2002:TOC}. Another example is the well known splitting \cite{Lorrain:1970:EFW} of the fixed frequency electromagnetic equations into
transverse electric waves, transverse magnetic waves, and transverse electromagnetic waves,
with the first two satisfying fixed frequency acoustic time harmonic type equations and the last satisfying a complex dielectric type equation.

For a large class of nonlinear equations
one has that
\beq \BJ_0=\BF(\BE_0)-\Bs_0,\quad \BGG_1\BE_0=\BE_0,\quad\BGG_1\BJ_0=0,
\eeq{nonlin}
where $\BF(\BE_0)$ is some nonlinear function of $\BE_0$, often monotone
in the sense that
\beq (\BF(\BE_1)-\BF(\BE_2),\BE_1-\BE_2)_{\CT}>0
\quad \text{for all}\quad \BE_1,\BE_2 \in \CE,\quad \text{with}\quad\BE_1\ne\BE_2,
\eeq{mon}
which ensures existence and uniqueness of solutions.

Considering a small perturbation of these fields, replacing $\BJ_0$, $\BE_0$,
and $\Bs_0$ with $\BJ_0+\BJ$, $\BE_0+\BE$, and $\Bs_0+\Bs$ at first order in these perturbations
we arrive back at the equations \eq{ad1} with $\BL(\Bx)$ being the ``tangent'' tensor $\BL=\Md \BF/\Md \BE|_{\BE=\BE_0}$.
An important example is the nonlinear conductivity equation (also called the generalized Poisson equation) which arises in a
number of very different physical systems \cite{Milgrom:2002:FNM}. We emphasize that in nonlinear conductivity the corresponding
``tangent'' tensor $\BL(\Bx)$ is not necessarily symmetric
and this breaking of symmetries occurs for other equations too, including  elasticity \cite{Ogden:1984:NED}.

In many cases of interest the equations \eq{nonlin} arise as the Euler-Lagrange equations associated with the minimization (assuming a minimum
exists) of
\beq \min_{\BE\in\CE}\int_{\mathbb{R}^3} [W(\BE(\Bx))-2\Bs_0(\Bx)\cdot\BE(\Bx)]\,d\Bx,
\eeq{nonlinw}
where $W(\BE(\Bx))$ is a nonlinear function (usually the free energy) that is often, but not necessarily, convex (instead quasiconvexity would suffice),  while
$\CE$ is the space of fields onto which $\BGG_1$ projects (so that $\BGG_1\BE=\BE$). Then the minimizer $\BE=\BE_0$ satisfies \eq{nonlin}
with $\BF(\BE_0)=\Md W(\BE)/\Md \BE|_{\BE=\BE_0}$. Small perturbations of the fields satisfy \eq{ad1} with $\BL=\Md^2 W(\BE)/\Md \BE^2|_{\BE=\BE_0}$
being symmetric.
\section{Periodic solutions in periodic media,
  and almost periodic solutions in periodic and nearly
  periodic media}
\setcounter{equation}{0}
\labsect{11}
If $\BL(\Bx)$, $\Bs(\Bx)$, $\BJ(\Bx)$, and $\BE(\Bx)$ are all periodic,
with unit cell $\GO$ then, in the absence of sources and with $\BL(\Bx)$ being constant,
the field equations are satisfied with constant fields $\BE$ and $\BJ$. So  $\BE$ and $\BJ$ no longer live in orthogonal spaces.
Rather one has to introduce the space $\CU$ of constant fields and the projection $\BGG_0$ onto it. Thus, we can generally write $\BJ$ and $\BE$
in the form
\beq \BJ=\BJ_0+\underline{\BJ}, \quad \BE=\BE_0+\underline{\BE}, \quad \BJ_0=\BGG_0\BJ,\quad \BE_0=\BGG_0\BE,
\eeq{per1}
where now $\underline{\BJ}$ and $\underline{\BE}$ live in
orthogonal spaces $\underline{\CJ}$ and $\underline{\CE}$, with the projections $\underline{\BGG}_1$ and  $\underline{\BGG}_2$ onto these
subspaces satisfying
\beq \BGG_0+\underline{\BGG}_1+\underline{\BGG}_2=\BI,\quad \BGG_0\underline{\BGG}_1=\BGG_0\underline{\BGG}_2=\underline{\BGG}_1\underline{\BGG}_2=0,
\eeq{per2}
where in Fourier space
\beqa \BGG_0(\Bk)&= & 0,\quad \underline{\BGG}_1(\Bk)=\BGG_1(\Bk), \quad  \underline{\BGG}_2(\Bk)=\BGG_2(\Bk),\quad\text{for}\quad \Bk\ne 0, \nonum
 \BGG_0(\Bk)&= & \BI,\quad \underline{\BGG}_1(\Bk)=0, \quad  \underline{\BGG}_2(\Bk)=0, \quad\text{for}\quad \Bk= 0.
 \eeqa{per3}
 Here $\Bk$ takes values on the discrete reciprocal lattice in Fourier space,
 discussed for example in Chapter 2 of \cite{Kittel:2005:ISS}.
 
Thus the equations now become
\beq \BJ=\BL\BE-\Bs, \quad \underline{\BGG}_1\BJ=0,\quad\underline{\BGG}_2\BE=0. \eeq{per4}
For a fixed source $\Bs$ and a given $\BE_0$ (that is called the applied
field) one solves the equations for $\underline{\BE}$, $\BJ_0$, and $\underline{\BJ}$. This is now directly a problem in the
theory of composites. Assuming a solution exists and is unique for every choice of $\BE_0$, one observes that $\BJ_0$ is linearly related
to $\BE_0$ and this linear relation
\beq \BJ_0=\BL_*\BE_0+\Bs_* \eeq{per5}
defines $\BL_*$ which is called the effective tensor, and $\Bs_*$ which may be called the effective source term. Here  $\BL_*$ can be obtained by solving
the equations for various values of $\BE_0$ with $\Bs=0$, while $\Bs_*$ can be obtained by solving the equations with $\BE_0=0$ (which does not imply $\BE=0$).
Typically, $\BL_*$ depends nonlinearly on $\BL$ and in a way that usually depends on the microstructure, and $\Bs_*$ also typically depends nonlinearly on
$\BL$ but linearly on $\Bs$, again in a way that usually depends on the microstructure (i.e., on the precise form of $\BL(\Bx)$ and $\Bs(\Bx)$ rather than say
just on their volume averages). As $\Bs_*$ depends linearly on $\Bs$ it suffices
to solve for the fundamental solutions (periodic Green's functions)
to calculate $\Bs_*$: these are the solutions with $\Bs(\Bx)$ being
an $\GO$-periodic arrays of delta functions. Since the solution
depends on the placement of the array with respect to $\BL(\Bx)$,
and cannot be obtained analytically, it is usually better to avoid
introducing these Green's functions if one is interested in $\Bs_*$
for a particular  $\Bs(\Bx)$. (On the other hand, periodic Green's functions
in homogeneous media are very helpful for obtaining $\BL_*$ for
simple geometries such as arrays of spheres:
see \cite{Borwein:2013:LST} and references therein).
In $N$-phase composites  $\Bs(\Bx)$
is often piecewise constant and then  $\Bs_*$ only
depends on $\Bs(\Bx)$ through its values  $\Bs_1, \Bs_2,\ldots,\Bs_n$
in the $N$ phases, and this relation is then governed by a response
tensor $\BCS_*$ dependent nonlinearly on $\BL(\Bx)$:
\beq \Bs_*=\BCS_*\BS, \eeq{linresp}
where $\BS=(\Bs_1, \Bs_2,\ldots,\Bs_n)$.
In the context on
thermoelasticity one can think of $\BCS_*$ as relating the effective
thermal expansion tensor to the thermal expansion tensors of the phases.
Such response tensors were introduced by Dvorak and Benveniste
\cite{Dvorak:1992:TSU} and the information contained in them
along with $\BL_*$  is embodied in the associated $\BY$-tensor
for multiphase materials: see Section 19.3
of \cite{Milton:2002:TOC}.

Effective tensors and effective sources play a pivotal role in the
theory of homogenization \cite{Bensoussan:1978:AAP, Zhikov:1994:HDO,
Milton:2002:TOC, Tartar:2009:GTH}. If the
the size of the unit cell of periodicity is small compared to macroscopic
variations in the fields, then often the macroscopic response is almost
the same as for
a homogeneous ``effective medium'' with constant tensor $\BL_*$ and a constant
supplementary source $\Bs_*$ that may be in addition to a macroscopic
source field $\Bs_m(\Bx)$. The macroscopic fields are
\beq \BE_m(\Bx)=\frac{1}{V(\GO)}\int_{\GO(\Bx)}\BE(\Bx')\,d\Bx', \quad
\BJ_m(\Bx)=\frac{1}{V(\GO)}\int_{\GO(\Bx)}\BJ(\Bx')\,d\Bx',\quad
\Bs_m(\Bx)=\frac{1}{V(\GO)}\int_{\GO(\Bx)}\Bs(\Bx')\,d\Bx',
\eeq{hom1}
where $\GO(\Bx)$ is the unit cell of periodicity, with volume $V(\GO)$,
shifted (by an amount not generally corresponding to a lattice vector)
so that its center at $\Bx$. Note that $\Bs_m(\Bx)$ washes out the
fluctuations in $\Bs(\Bx')$ within each period cell and these fluctuations
along with $\Bs_m(\Bx)$ and $\BL(\Bx)$,
are what determines $\Bs_*$. We define the fluctuating component $\Bs_f(\Bx)$
of $\Bs(\Bx)$ as
\beq \Bs_f(\Bx)=\Bs(\Bx)-\Bs_m(\Bx), \eeq{hom1a}
and we assume this is almost periodic. 

The macroscopic fields approximately satisfy
\beq \BE_m(\Bx)=\BL_*(\Bx)\BJ_m(\Bx)-\Bs_*(\Bx),\quad
\BGG_1(\Bk)\BE_m=\BE_m,\quad \BGG_1(\Bk)\BJ_m=0 \eeq{hom.1}
where we have allowed $\BL_*$ and $\Bs_*$ to depend on $\Bx$, on the macroscopic
scale, as $\BL(\Bx)$ and $\Bs_f(\Bx)$ may not be quite periodic. Not quite
periodic functions $\BL(\Bx)$ and $\Bs(\Bx)$ can be obtained from
functions $\underline{\BL}(\Bx,\By)$ and $\underline{\Bs}(\Bx,\By)$
that are periodic in $\By$ (called the fast variable) but are smooth
functions of $\Bx$ (the slow variable). Then by setting
\beq \BL(\Bx)=\underline{\BL}(\Bx,\Bx/\Ge), \quad
\Bs(\Bx)=\underline{\Bs}(\Bx,\Bx/\Ge)
\eeq{hom.1a}
where $\Ge$ is a small parameter, one obtains almost periodic functions
$\BL(\Bx)$ and $\Bs_f(\Bx)$ \cite{Bensoussan:1978:AAP}. For small $\Ge$, the
resulting fields $\BE(\Bx)$ and $\BJ(\Bx)$ are also in correspondence
with two variable functions $\underline{\BE}(\Bx,\By)$ and $\underline{\BJ}(\Bx,\By)$:
\beq  \BE(\Bx)\approx\underline{\BE}(\Bx,\Bx/\Ge),\quad
\BJ(\Bx)\approx\underline{\BJ}(\Bx,\Bx/\Ge).
\eeq{hom.1b}
This is the basis of two scale homogenization \cite{Nguetseng:1989:GCR, Allaire:1992:HTS}.

From the definitions \eq{hom1} it follows that
the macroscopic fields $\BE_m(\Bx)$ and $\BJ_m(\Bx)$ satisfy exactly
the same differential constraints as the local fields
$\BE(\Bx)$ and $\BJ(\Bx)$. However, to determine $\BL_*$ and $\Bs_*$
it may suffice to work with a set of uncoupled equations each
having a $\widetilde{\BGG}^{(j)}_1(\Bk)$, indexed by $j$,
associated with different powers of $\Bk$ in $\BGG_1(\Bk)$.
These  $\widetilde{\BGG}^{(j)}_1(\Bk)$ satisfy
$\widetilde{\BGG}_1^{(j)}(\Bk)=\widetilde{\BGG}_1^{(j)}(\Bk/|\Bk|)$.
The reason for the uncoupling is that there is a separation of length scales
when one is interested in taking the homogenization
limit as the size of the unit cell tends to zero.
Examples are the time harmonic acoustic, electromagnetic, and
elastic wave equations (for which the $\BGG_1(\Bk)$ will be introduced
in the next section). Then the $\widetilde{\BGG}_1^{(j)}(\Bk/|\Bk|)$
are those associated with the corresponding quasistatic equations,
and solving these in the periodic setting enables one  
to determine  $\BL_*$ and $\Bs_*$.

Although rarely discussed, effective source
terms are relevant whenever one has sources $\Bs(\Bx)$ with the same periodicity
as $\BL(\Bx)$. An example of where effective sources arise is in
elastostatics with thermal expansion as in
\eq{15.4}. Then $\Bs(\Bx)=\Gt\BGa(\Bx)$, where $\Gt$ is the constant temperature rise and $\BGa(\Bx)$ is the thermal expansion tensor field,
and $\Bs_*=\Gt\BGa_*$ where $\BGa_*$ is the effective tensor of thermal expansion. For isotropic composites of two isotropic components
having bulk moduli $\Gk_1$, $\Gk_2$ and thermal expansion tensors $\Ga_1$ and $\Ga_2$ there is Levin's formula \cite{Levin:1967:TEC}:
\beq \Ga_*=\frac{\Ga_1(1/\Gk_*-1/\Gk_2)-\Ga_2(1/\Gk_*-1/\Gk_1)}{1/\Gk_1-1/\Gk_2},
\eeq{per6}
that links the effective thermal expansion tensor $\Ga_*\BI$ to the effective bulk modulus $\Gk_*$. It can be explained by the observing that with the right
combination of temperature change and pressure increase, both phases expand at
exactly the same rate, and hence the composite must expand at this rate.

This formula is an example of an exact relation, an identity that is independent of the microstructure. More generally an exact relation is a manifold $\CM$ of
supertensors such that $\BL_*\in\CM$ whenever $\BL(\Bx)\in\CM$ for all $\Bx$ (assuming $\BL(\Bx)$ some constraints that ensure $\BL_*$ exists)
\cite{Grabovsky:1998:EREa}.
In the context of Levin's result, $\BL$ is the tensor entering \eq{15.5}. 
For a long while exact relations were obtained one at a time by dozens of researchers, some quite famous (see, for example, Chapters 3, 4,
5, and 6 in \cite{Milton:2002:TOC}). Grabovsky and
Sage \cite{Grabovsky:1998:EREa} launched the unifying theory of exact
relations by identifying manifolds $\CM_L$
such that $\BL_*\in\CM_L$ whenever $\BL(\Bx)\in\CM_L$ for all $\Bx$, under the
restriction that $\BL(\Bx)$ is in the class of
hierarchical laminate geometries
(obtained by laminating tensors in $\CM_L$ in different directions
on well separated length scales). This then led to a general theory
that guaranteed an exact relation would hold for all
geometries \cite{Grabovsky:2000:ERE} not just laminate ones. An example
of a relation which holds for laminate geometries but not more general ones
was discovered by Grabovsky \cite{Grabovsky:2017:MIF}.
The theory is reviewed, for example, in Chapter 17 of \cite{Milton:2002:TOC}, the article \cite{Grabovsky:2004:AGC},
and more comprehensively in the book \cite{Grabovsky:2016:CMM}.
When it boils down to it, the theory of exact relations consists of identifying subspaces $\CK$ in the space of supertensors with the algebraic
property that
\beq \BK_1\BGG(\Bk)\BK_2\in\CK,\quad\text{for all }\Bk\ne 0,\quad\text{and for all }\BK_1,\BK_2\in\CK,
\eeq{per7}
where
\beq \BGG(\Bk)=\BGG_1(\Bk)[\BGG_1\BL_0\BGG_1]^{-1}\BGG_1,
\eeq{per8}
in which the inverse is to be taken on the space on which $\BGG_1(\Bk)$ projects, and $\BL_0\in\CM$. The relation between $\CK$ and $\CM$ is that
$\BL\in\CM$ if and only if
\beq \BK=(\BL-\BL_0)[\BI+\BGG(\Bk_0)(\BL-\BL_0)]^{-1} \in \CK.
\eeq{per9}
This relation between $\CK$ and $\CM$ is independent of the choices of $\BL_0\in\CM$ and $\Bk_0\ne 0$.
This systematized the study of exact relations, and resulted in a flood of
new ones derived by Yury Grabovsky and collaborators,
dwarfing the total number previously obtained,
many listed in \cite{Grabovsky:2016:CMM}. So far studies have been confined to manifolds $\CM$ that have suitable
rotational invariance properties so that \eq{per7} only needs to be checked for one value of $\Bk$.

Falling under the umbrella of exact relations are links where, for example, one establishes relations between the effective tensors $\BL_*^{(1)}$ and $\BL_*^{(2)}$ of two different
problems, that could even be physically unlike each other. In this setting it suffices to look for manifolds where $\BL$ has two blocks along its diagonal,
corresponding to the two problems we hope to link, thus with $\BL_*$ having 
$\BL_*^{(1)}$ and $\BL_*^{(2)}$ as blocks on its diagonal.

The theory of exact relations for composites has more recently led to
universal (geometry independent) exact identities satisfied by the infinite body
Green's function for inhomogeneous media when the tensor field $\BL(\Bx)$
(no longer restricted to be periodic) takes values in $\CM$,
i.e. $\BL(\Bx)\in\CM$ for all $\Bx$. These lead to exact relations
satisfied by the Dirichlet-to-Neumann map of bodies when $\BL(\Bx)\in\CM$
inside the body, and to a flood
of new conservation laws called boundary
field equalities \cite{Milton:2019:ERG}, when one has suitable boundary
conditions. Note that $\BL(\Bx)\in\CM$ can just be viewed as a constraint
on the fields inside the body, and then this constraint is independent
of $\Bx$.

\section{Some general results that apply to the
  effective tensor}
\setcounter{equation}{0}
\labsect{12}

Lets show that if $\BL$ is replaced by its adjoint $\BL^\dagger$, then
the effective tensor $\BL_*$ is replaced by $(\BL^\dagger)_*=(\BL_*)^\dagger$.
Here the adjoint is taken with respect to the inner
product given by \eq{innp} where the integral is taken over the unit cell of periodicity. Consider, with $\Bs=0$, one solution to \eq{per4}
and a second solution to the adjoint equation:
\beq \BJ^{(1)}=\BL\BE^{(1)}, \quad  \BJ^{(2)}=\BL^\dagger\BE^{(2)},
\quad \underline{\BGG}^{(1)}\BJ^{(1)}=\underline{\BGG}^{(1)}\BJ^{(2)}=0,\quad
\underline{\BGG}^{(2)}\BE^{(1)}=  \underline{\BGG}^{(2)}\BE^{(2)}=0.
\eeq{p10}
Using the orthogonality of the subspaces $\CU$, $\underline{\CE}$
and $\underline{\CJ}$ we see that 
\beqa (\BE^{(1)}_0,(\BL_*)^\dagger\BE^{(2)}_0)_\CT & = & (\BL_*\BE^{(1)}_0,\BE^{(2)}_0)_\CT = (\BJ^{(1)}_0,\BE^{(2)}_0)_\CT=(\BJ^{(1)},\BE^{(2)})=
(\BL\BE^{(1)},\BE^{(2)})=(\BE^{(1)},\BL^\dagger\BE^{(2)}) \nonum
& = & (\BE^{(1)},\BJ^{(2)})=(\BE^{(1)}_0,\BJ^{(2)}_0)_\CT
=(\BE^{(1)}_0,(\BL^\dagger)_*\BE^{(2)}_0)_\CT,
\eeqa{p11}
which implies $(\BL^\dagger)_*=(\BL_*)^\dagger$. In particular, if $\BL$ is self
adjoint then so is $\BL_*$. 

A simple formula can be obtained for the change in the effective tensor
$\BL_*$ when the tensor field $\BL(\Bx)$ is perturbed slightly.
Suppose $\BL$ depends on some parameter $\Gn$, $\BL=\BL(\Gn)$, while
$\BE_0^{(1)}$ and $\BE_0^{(2)}$ do not. Then following Section
16.1 in \cite{Milton:2002:TOC} we have
\beqa \left(\left[\frac{d}{d \Gn}(\BL_*(\Gn))\right]\BE_0^{(1)},\BE_0^{(2)}\right)_\CT
& = & \frac{d}{d \Gn}(\BL_*(\Gn)\BE_0^{(1)},\BE_0^{(2)})_\CT
= \frac{d}{d \Gn}(\BL(\Gn)\BE^{(1)}(\Gn),\BE^{(2)}(\Gn))\nonum
&= &\left(\frac{d\BL(\Gn)}{d\Gn}\BE^{(1)}(\Gn), \BE^{(2)}(\Gn)\right)
+\left(\frac{d\BE^{(1)}(\Gn)}{d\Gn},\BL^\dagger(\Gn)\BE^{(2)}(\Gn)\right)
+\left(\BL(\Gn)\BE_1(\Gn),\frac{d\BE^{(2)}(\Gn)}{d\Gn}\right). \nonum &~&
\eeqa{p12}
Since $d\BE^{(1)}(\Gn)/d\Gn$ and  $d\BE^{(2)}(\Gn)/d\Gn$
both lie in the space $\underline{\CE}$, while $\BJ^{(2)}=\BL^\dagger(\Gn)\BE^{(2)}(\Gn)$
and $\BJ^{(1)}=\BL(\Gn)\BE^{(1)}(\Gn)$ lie in $\CU\oplus\underline{\CJ}$,
it follows that the last two terms in \eq{p12} are zero and we have
\beq \left(\left[\frac{d}{d \Gn}(\BL_*(\Gn))\right]\BE_0^{(1)},\BE_0^{(2)}\right)_\CT
=\left(\frac{d\BL(\Gn)}{d\Gn}\BE^{(1)}(\Gn), \BE^{(2)}(\Gn)\right).
\eeq{p13}
In other words, if $\BL$ is perturbed to $\BL+\Ge\BL$ then to
first order in $\Ge$, $\BL_*$ will get perturbed to $\BL_*+\Ge\BL_*'$
where
\beq
(\BL_*' \BE_0^{(1)},\BE_0^{(2)})_\CT=(\BL' \BE^{(1)},\BE^{(2)}).
\eeq{p14}
In other words, one can calculate $\BL_*'$ just from the fields that solve
the unperturbed problem. In particular, this perturbation could be due to a
small magnetic field, giving a formula for the Hall coefficient in terms of the
fields solving the conductivity equations
with no magnetic field present \cite{Bergman:1983:SDL, Briane:2009:HTD}. 
This was used to show that in certain geometries of interlinked rings,
suggested by chain mail artist Dylon Whyte, the sign of the Hall
coefficient could be reversed
from the sign of the Hall coefficient
of the constituent materials \cite{Briane:2009:HTD}.
It disproved the common perception that the sign of the Hall coefficient
determines the sign of the charge carriers. The
perception was based on the model where the electrons travel in straight
lines, which is certainly not the case in these microgeometries.
Later the design was considerably
simplified \cite{Kadic:2015:HES} and the sign change of the Hall coefficient demonstrated
in experiments with the interlinked ring geometries, that had a semiconductor surface coating, amazingly
replicated using three dimensional laser lithography \cite{Kern:2016:EES}.  
In suitable geometries one can also get novel effects such as the parallel
Hall effected where the electric field induced by the magnetic field
are both parallel rather than perpendicular \cite{Briane:2010:AEH, Kern:2017:EPH}.
The perturbation analysis also applies to piezoelectric, thermoelectricity,
and any other coupled equation where the coupling is weak, and can
be used in an inverse fashion  to determine or bound
the variance of the field $\BE$: see Sections 16.2 and 16.4 of
\cite{Milton:2002:TOC} and references therein.

\section*{Acknowledgements}
GWM thanks the National Science Foundation for support through grant DMS-1814854, and Christian Kern for helpful comments on the manuscript, and
for providing additional references. The work was largely based on the
books \cite{Milton:2002:TOC, Milton:2016:ETC} and in the context of the latter book I would like to thank Nelson Beebe for his enormous help in preparing it, and 
Maxence Cassier, Kirill Cherednichenko, Elena Cherkaev, Richard Craster, Vikram Gavani, Davit Harutyunyan, Michael Fisher,
Richard James, Hyeonbae Kang, Paul Martin, Ornella Mattei, Mordehai  Milgrom,  Alexander Movchan, Mihai Putinar, Pierre Seppecher,
Ping Sheng, Fernando Guevara Vasquez, Martin Wegener,
Aaron Welters, and John Willis for their helpful feedback on it. I am grateful to Yury Grabovsky and Pradeep Sharma for their reviews of it.
Those books and the present paper were
heavily influenced and propelled by the ideas of many mentors, colleagues, and friends, particularly Andrej Cherkaev, Yury Grabovsky,
Ross McPhedran, Alexander Movchan, George Papanicolaou, Pierre Seppecher, and  John Willis. Aaron Welters and Christian Kern are thanked for
bringing the Faraday effect to the attention of the author.

\ifx \bblindex \undefined \def \bblindex #1{} \fi\ifx \bbljournal \undefined
  \def \bbljournal #1{{\em #1}\index{#1@{\em #1}}} \fi\ifx \bblnumber
  \undefined \def \bblnumber #1{{\bf #1}} \fi\ifx \bblvolume \undefined \def
  \bblvolume #1{{\bf #1}} \fi\ifx \noopsort \undefined \def \noopsort #1{}
  \fi\ifx \bblindex \undefined \def \bblindex #1{} \fi\ifx \bbljournal
  \undefined \def \bbljournal #1{{\em #1}\index{#1@{\em #1}}} \fi\ifx
  \bblnumber \undefined \def \bblnumber #1{{\bf #1}} \fi\ifx \bblvolume
  \undefined \def \bblvolume #1{{\bf #1}} \fi\ifx \noopsort \undefined \def
  \noopsort #1{} \fi

\end{document}